\documentclass[12pt]{article}
\usepackage[T1]{fontenc}
\usepackage{babel}
\usepackage{refstyle}
\usepackage{amsmath}
\usepackage{amssymb}
\usepackage{setspace}
\usepackage[numbers]{natbib}
\makeatletter

\AtBeginDocument{\providecommand\eqref[1]{\ref{eq:#1}}}
\AtBeginDocument{}
\AtBeginDocument{\providecommand\Eqref[1]{\ref{Eq:#1}}}
\RS@ifundefined{subsecref}
  {\newref{subsec}{name = \RSsectxt}}
  {}
\RS@ifundefined{thmref}
  {\def\RSthmtxt{theorem~}\newref{thm}{name = \RSthmtxt}}
  {}
\RS@ifundefined{lemref}
  {\def\RSlemtxt{lemma~}\newref{lem}{name = \RSlemtxt}}
  {}

\numberwithin{equation}{section}

\@ifundefined{date}{}{\date{}}
\makeatother

\newcommand{\be}{\begin{equation}}
\newcommand{\ee}{\end{equation}}
\newcommand{\bea}{\begin{eqnarray}}
\newcommand{\eea}{\end{eqnarray}}

\def\({\left(} \def\){\right)}

\begin{document}
\title{\vspace{-1.8in}
{Defrosting frozen stars: \\ spectrum of  non-radial oscillations}}
\author{\large Ram Brustein${}^{(1)}$,  A.J.M. Medved${}^{(2,3)}$, Tom Shindelman${}^{(1)}$
\\
\vspace{-.5in} \hspace{-1.5in} \vbox{
\begin{flushleft}
 $^{\textrm{\normalsize
(1)\ Department of Physics, Ben-Gurion University,
   Beer-Sheva 84105, Israel}}$
$^{\textrm{\normalsize (2)\ Department of Physics \& Electronics, Rhodes University,
 Grahamstown 6140, South Africa}}$
$^{\textrm{\normalsize (3)\ National Institute for Theoretical Physics (NITheP), Western Cape 7602,
South Africa}}$
\\ \small \hspace{0.57in}
   ramyb@bgu.ac.il,\  j.medved@ru.ac.za,\ tomshin@post.bgu.ac.il
\end{flushleft}
}}
\date{}
\maketitle

\renewcommand{\baselinestretch}{1.15}

\begin{abstract}

The frozen star model describes a type of black hole mimicker; that is, a regular, horizonless, ultracompact object that behaves just like a Schwarzschild black hole from an external-observer's perspective. In particular, the frozen star is bald, meaning that it cannot be excited.  To mimic the possible excitations of the frozen star, it needs to be ``defrosted''  by allowing deviations from the maximally negative radial pressure and vanishing tangential pressure of the fluid sourcing the star.  Here, we extend a previous study on non-radial oscillations of the defrosted star by considering, in addition to the fluid modes, the even-parity metric perturbations and their coupling to the fluid modes. At first, general equations are obtained  for the perturbations of  the energy density and  pressure along with the even-parity perturbations of the metric for a static, spherically symmetric but otherwise generic background with an anisotropic fluid. This formal framework is then applied to the case of  a defrosted star. The spectrum of non-radial oscillations is obtained to leading order in an expansion in terms of $\gamma$, which is the small relative deviation away from maximally negative radial pressure. We find that the sound velocity of the modes is non-relativistic, and proportional to $\gamma$, while their lifetime is parametrically long, proportional to $1/\gamma^2$. This result was anticipated by previous discussions on the collapsed polymer model, whose strongly non-classical interior is argued to provide a microscopic description of the frozen and defrosted star geometries. Our results will serve as a starting point for calculating the spectrum of emitted gravitational waves from an excited frozen star.
\end{abstract}
\maketitle

\newpage
\renewcommand{\baselinestretch}{1.5}

\section{Introduction}

The ``frozen star'' is a particular class of black hole (BH) mimickers,
by which we mean regular, horizonless, ultracompact  objects that, from
the perspective of an externally situated observer, behave just like
the BHs of general relativity, including their standard semiclassical aspects.
The original discussions on the frozen star model, although not yet called as such,  can be found in
 \cite{bookdill,BHfollies}. Subsequent treatments that do indeed incorporate the name
 frozen star,  which was adopted as an homage to early literature \cite{Ruffini:1971bza},
 can be found in   \cite{popstar,trajectory,fluctuations,U4Euclidean}. Even more recent  discussions  have allowed for the incorporation of rotation \cite{NotSteveKerr}  and provided a formal description of the matter source in terms of a fluid of electric-flux tubes \cite{FluxUstat}.

 One of the signature features of the frozen star model
 is a radial pressure   that is  maximally negative,  $\;p\equiv p_r= -\rho\;$, $\rho$ being
 the energy density, and  a vanishing pressure  in the tangential directions  $\;q \equiv  p_{\theta}=p_{\phi}=0\;$.
 The former  property allows the model to evade the singularity theorems \cite{PenHawk1,PenHawk2} and Buchdahl-like compactness-of-matter bounds \cite{Buchdahl,chand1,chand2,bondi,MM}, while still  respecting the null-energy condition,  whereas the latter is a consequence of the microscopic
 description of the star's interior being effectively two dimensional \cite{strungout}. Another important feature is that the frozen star solution deviates
 from the Schwarzschild geometry over
 horizon-sized length scales, meaning that it is not plagued by  conflicts
 with the conservation of energy that other ultracompact objects have while slowly evaporating
\cite{frolov,visser}.

Yet another prominent feature is that the  frozen star metric is nearly null
throughout the interior, as each radial surface is timelike up to an extremely  small, dimensionless parameter $\varepsilon^2$. As explained in detail elsewhere \cite{bookdill,BHfollies} and vindicated in \cite{U4Euclidean}, this strange property is  just what is needed to account for an area-law's   worth of entropy  without having a formal horizon. As far as we know, the frozen star may be the only model of its kind that can reproduce all of the standard properties of a BH --- including  all of  its thermodynamic properties  --- but with none of the usual pathologies. For an  incomplete list of other models, see
\cite{eg1,eg2,eg3,eg4,eg5,eg6,eg7,eg8,eg9,eg10,eg11}. Further references
can be found in the review articles \cite{egr1,egr2,egr3,carded}.

It is worthwhile to recall that the frozen star's  microscopic description, what is known as
the collapsed polymer model \cite{inny,strungout,emerge}, is provided by a fluid of  highly excited, interacting,
long, closed, fundamental strings. In this sense, the frozen star metric  is meant to represent how an external observer would effectively  describe  the interior geometry, even though the inside  is lacking a classically geometrical description at a fundamental level. From this perspective, it would be  inaccurate to claim that astrophysical BHs are filled
with some highly anisotropic classical fluid. The importance of the metric is
then not to paint a geometrical portrait but to  enable  one to perform precise calculations within the realms of general relativity and quantum field theory in curved space.
Nevertheless, a recent discovery \cite{FluxUstat} suggests that a particular
arrangement of open strings, or lines of electric flux,  that are radially directed from the
center of the star to its outer surface (or {\em vice versa}) can act as the matter source for the frozen star geometry. This description of the interior also has its origins in string theory
\cite{GHY,Yi} but, unlike the polymer model,  is directly endowed with a
well-defined metric via Einstein's equations. It is also worth mentioning that this radial-string picture
is consistent with a  hedgehog-like description of the frozen star interior, as presented in \cite{trajectory} and  originally discussed
in \cite{hedge1,hedge2}.

A direct consequence of the frozen star equation of state $\;p+\rho=0\;$  is the ultrastability of the model; any perturbations of the interior geometry or matter densities
vanish identically  \cite{bookdill,popstar}. This feature agrees with the polymer model, which is completely stable, behaving like a hairless BH, in the absence of quantum effects or, equivalently, as the closed-string coupling goes to zero, $\;g_s^2\to 0\;$. When quantum effects are included, it was shown that the polymer  supports a spectrum of non-radial oscillations whose sound velocity scales as $g_s$ (in $\;c=1\;$ units) and whose lifetime scales as $1/g_s^2$ \cite{collision}.
So that,  to mimic the quantum effects of the polymer model and study the spectrum of  oscillatory modes of the frozen star, one must allow for deviations away from its maximally negative radial pressure by, effectively, ``defrosting'' the star. For the calculations to make sense, such deviations --- which will be  denoted by the dimensionless, non-negative parameter $\gamma$  ---  need to be small enough to be regarded as perturbative, but not so extremely small  that $\gamma$ dominates over $\varepsilon^2$. The consequences of these constraints will be a point of emphasis in Section~2. What is already clear though is that $\gamma^2$ plays the same role as $g_s^2$ does in the polymer model; as demonstrated explicitly in Eqs.~(\ref{eq:gttgamma}) and~(\ref{grr}), which inform us that radial  velocities scale as $\gamma$.

The model of a frozen star with a relaxed equation of state has already been introduced in \cite{fluctuations}, where we determined the nature of
the non-radial oscillatory modes of the fluid perturbations in the energy density, radial pressure and tangential pressure: $\delta \rho$, $\delta p$ and $\delta q$, respectively. In \cite{fluctuations}, following \cite{showerheads}, we applied the Cowling approximation \cite{cowling}, meaning that the  metric perturbations are ignored completely. This is a reasonable approximation for BH mimickers, as it becomes more accurate as the object becomes more compact \cite{KSmodel}. Nevertheless, a central goal of the current paper is to relax this approximation and thus  allow for coupling between the fluid and geometric sectors.

Another objective of the the current paper, which was not pursued  in \cite{fluctuations}, is to solve for the spectrum of the ringdown modes, as this can only be obtained by coupling oscillatory modes to the external spacetime; in which case, the gravitational perturbations can no longer be dismissed. The ringdown modes are of great interest because these determine the spectrum of  gravitational waves (GWs) that are emitted  from an excited star; for example, in the final phase of a merger event in a binary system.
Then,  if astrophysical BHs are indeed described by frozen stars, their emission of GWs could  possibly lead to observable predictions ({\em e.g.}, \cite{LIGO,LIGO1}).  Put differently, the ringdown modes provide a means for probing the star's interior.

In this regard, a word of caution is in order: It is our expectation that a BH mimicker of the frozen star class would, for all practical purposes,  be indistinguishable from a Schwarzschild BH of the same mass when it is in its equilibrium state \cite{ridethewave}. If this is strictly true, what will be required to discriminate between our model and the Schwarzschild solution, as well as between frozen stars and other candidate models, is precise data from BH mergers that is capable of  probing their spectrum of excitations.

As the frozen star and polymer models are suppose to be complementary descriptions of
the same object, an important consistency check would be if they produced
similar spectra for their respective ringdown modes.
What is then pertinent is the aforementioned article \cite{collision} (also
see \cite{spinny}) in which  the mode spectrum for the polymer model was evaluated.~\footnote{Albeit with some heuristic inputs due to the lack of an interior geometry.}  In the present article, we show that the two spectra are indeed very similar, given that
$\;\gamma^2$ is identified  as the star's counterpart to  the polymer's $g_s^2$.

Although much of the discussion in the current
work revolves around the particular choice of model being the frozen (or, rather, defrosted) star, the equations  presented below are applicable to any anisotropic stellar model,  given the specification of an equation of state and suitable boundary conditions. Our generic analysis is supplemented by a detailed appendix, in which our derivations are shown to differ from previous attempts at this sort of description
\cite{more0,more1,more2,more3,more4,mondal2023nonradial}.
In most cases, this can be attributed to the specificity of the analysis;
for instance, \cite{more4} considers the particular case of polytropes.
On the other hand, in \cite{mondal2023nonradial},  more general
calculations are performed; however, the authors make a restricting assumption that the perturbations respect the spherical symmetry of the background.~\footnote{More specifically,
they set the perturbations $\delta s_{\theta}$  and $\delta s_{\phi}$ to be
{\em a priori} vanishing,  where the vector $s_{\mu}$ is the unit normal to the fluid velocity as expressed in the notation of \cite{mondal2023nonradial} (we rather use $k_{\mu}$).}

The rest of the paper is arranged as follows:
We start by introducing the relevant background solutions; those of a  spherically symmetric, static but otherwise generic
anisotropic star (Sec.~2.1) and the defrosted star model (Sec.~2.2).
The second  section ends with a discussion on the pair of dimensionless, perturbative parameters, $\varepsilon^2$ and $\gamma^2$,
that are needed to describe the defrosted star geometry (Sec.~2.3).
The perturbative analysis is then carried out; first for the generic background
(Sec.~3), with many of the details deferred to the Appendix.
The generic results are then applied to the defrosted star (Sec.~4), culminating with a derivation of the spectrum for the ringdown modes (Sec.~4.3). The
main part of the paper
ends with a discussion (Sec.~5).

\section{Background geometry}

\subsection{A generic anisotropic star}

The perturbation equations of an anisotropic star will eventually be  derived in Section~3
and the Appendix. For now, we begin with the background solution, which
is static, spherically symmetric but otherwise generic,
\begin{equation}
ds^{2}\;=\;-e^{2\Phi}dt^{2}+e^{2\Lambda}dr^{2}+r^{2}d\Omega^{2}\;, \label{eq:backgroundmetric}
\end{equation}
where $\;\Phi=\Phi(r)\;$ and $\;\Lambda=\Lambda(r)\;$.

The advertised  anisotropy enters through the energy--momentum (EM) tensor
\begin{equation}
T_{\mu\nu}\;=\;\rho u_{\mu}u_{\nu}+pk_{\mu}k_{\nu}+q\left(g_{\mu\nu}+u_{\mu}u_{\nu}-k_{\mu}k_{\nu}\right)\;,\label{eq:SET}
\end{equation}
where the tangential pressure $q$ is allowed to differ from the radial
pressure $p$,  $u^{\mu}$ is the fluid 4-velocity and $k^{\mu}$
is a radial unit vector, $\;k^{\mu}k_{\mu}=+1\;$,
that satisfies $\;u^{\mu}k_{\mu}=0\;$.

We  will denote metric perturbations by $h_{\mu\nu}$ so that the components of
the metric can be written as
\begin{equation}
g_{\mu\nu}\;=\;g_{\mu\nu}^{(0)}+h_{\mu\nu}\;,
\end{equation}
where the  $g_{\mu\nu}^{(0)}$ components are  defined in Eq.~(\ref{eq:backgroundmetric}).

The unperturbed
Einstein equations are (a prime denotes a radial derivative)
\begin{align}
\left(re^{-2\Lambda}\right)^{\prime} &\; =\;1-8\pi G \; r^{2}\rho\;,\label{eq:rho}\\
e^{-2\left(\Lambda+\Phi\right)}\left(re^{2\Phi}\right)^{\prime} & \;=\;1+8\pi G \; r^{2}p\;,\label{eq:p}\\
\frac{dp}{dr} & \;=\;-\Phi^{\prime}\left(p+\rho\right)-\frac{2}{r}\left(p-q\right)\;.\label{eq:p'}
\end{align}

Now, supplemented by a mass function,
\begin{equation}
m\left(r\right)\;=\;4\pi \int_{0}^{r}\rho\left(s\right)s^{2}\text{d}s\;,
\end{equation}
Eq.~(\ref{eq:rho}) translates into
\begin{equation}
e^{-2\Lambda}\;=\;1-2 G m\left(r\right)/r\;,
\end{equation}
while  Eq.~(\ref{eq:p}) adopts  the form
\begin{equation}
1+2r\frac{d\Phi}{dr}\;=\;\frac{1+8\pi r^{2}p}{1-2Gm\left(r\right)/r}\;.
\end{equation}
Combining the previous equation  with Eq.~(\ref{eq:p'}), one obtains a   modified
version of the standard Tolman--Oppenheimer--Volkoff equation,
\begin{equation}
\frac{dp}{dr}\;=\;-G\frac{\left(p+\rho\right)\left(m\left(r\right)+4\pi  r^{3}p\right)}{r\left(r-2G m\left(r\right)\right)}-\frac{2}{r}\left(p-q\right)\;.
\end{equation}

Ultimately, in order to complete the full perturbative analysis as described below,
an equation of state of the form $\;\rho=\rho\left(p\right)\;$, $\;q=q\left(p\right)\;$ would have to be specified.

\subsection{The defrosted star model}

In this case, the relevant metric components are
\bea
-g_{tt}&=& e^{2\Phi}\;=\; \varepsilon^2+ \gamma\left(\frac{r}{R}\right)^a\;, \\
g^{rr}&=&  e^{2\Lambda} \;=\; \varepsilon^2 +\gamma\left(\frac{r}{R}\right)^b\;,
\eea
where all parameters are dimensionless besides $R$ (the radius of the star) and the original  frozen star model is retrieved by setting
$\;\gamma=0\;$.
Although $a$ and $b$ are {\em a priori} undetermined numbers, it was shown in \cite{fluctuations} that self-consistency requires the specific choices of
$\;a=2\;$ and $\;b=0\;$, so that
\bea
-g_{tt}&=& \gamma\left(\frac{r}{R}\right)^2\;, \label{eq:gttgamma} \\
g^{rr}&=& \gamma\;, \label{grr}
\eea
with $\varepsilon^2$ now neglected (both here and in what follows) because  $\;\varepsilon\ll \gamma\;$. Subsequent expressions will
also be to linear order in $\gamma$  because $\;\gamma\ll 1\;$.
This hierarchy  of scales is discussed at length in Section~2.3.

It should be noted that the form of the metric is altered near the
center of the star \cite{trajectory}
and close to its outer surface \cite{popstar}. However, the former has no bearing on the current analysis and similarly for the latter given that the thin-wall approximation has been implemented, as we choose to do here.

Let us now consider two components  of the EM tensor,
\be
8\pi G r^2 \rho \;=\;1-\gamma
\;,
\label{r2r}
\ee
and
\be
8\pi G r^2  p \;=\;-1+3\gamma
\;,
\label{r2p}
\ee
which can be combined into
\be
p\;=\;\rho \left(-1+2\gamma\right)\;
\label{pvr}
\ee
or
\be
8\pi G r^2  (\rho+p)\;=\; 2\gamma\;,
\label{r2rplusp}
\ee
which makes  the intended deviation from $\;\rho+p=0\;$ quite evident.
Also note that $\;\rho+p \geq 0\;$, as required by causality.

Via the background conservation equation for the EM tensor,
\begin{equation}
\frac{dp}{dr}\;=\;-\partial_{r}\Phi\left(\rho+p\right)-\frac{(p-q)}{r}\;,
\label{energycons}
\end{equation}
one finds that the tangential pressure $q$ is no longer vanishing, as it is
in the frozen star model. Rather,
\be
8 \pi G  r^2 q \;=\; \gamma  \;
 \label{r2q}
\ee
or
\be
  q \;=\; \frac{1}{2}\left(\rho+p\right)\;,  \;
 \label{eq:q}
 \ee
from which it can be deduced that (also using  Eq.~(\ref{pvr}))
\bea
\frac{\partial q}{\partial p}&=&-\gamma   \;.
 \label{dpdq}
\eea

Another useful relation is
\be
\partial_r \Phi\; =\;\frac{2}{r}\;.
\label{drphi}
\ee

The radius $R$  of a defrosted  star of mass $M$ is larger than its Schwarzschild size, as can be shown
by matching the radial component of the defrosted star metric to the  standard Schwarzschild
form at $\;r=R\;$. This leads to, at linear order in $\gamma$,
$\;R=2GM(1+\gamma)\;$. That the
star's mass is indeed $M$ (again at linear order)  follows from
\be
\int\limits^{R}_{0} dr\; 4\pi  r^2 \; \rho(r)\; = \;\frac{1}{2G}\int\limits^{R}_{0} dr\left(1-\gamma\right)
\;=\; \frac{1-\gamma}{2G}R \;=\; M\;,
\ee
where $\rho(r)$  has been  obtained from Eq.~(\ref{r2r}).
Since the defrosted star is meant to be regarded as an excited state of the frozen star \cite{collision},
the implication is that the ground state had an initial mass of less than $M$.

\subsection{Interpretation of $\varepsilon^2$ and $\gamma^2$ }

To gain some insight into the defrosted star model, it is useful to discuss the interpretation and relative size of the model's pair of small, dimensionless   parameters $\varepsilon^2$ and $\gamma^2$. Although we cannot be precise on either, our working assumption is that these two parameters correspond to the two small parameters of the polymer model: $\varepsilon^2$ corresponds to $\;\epsilon=\frac{l_P}{R}$, the Planck  length measured in Schwarzschild units, and $\gamma^2$ corresponds to  $g_s^2\;$,  the closed-string coupling.  It is necessary for the self-consistency of the polymer model that $\;\epsilon\ll g_s^2\;$. This would certainly be true for any observable BH given that $g_s^2$ is not abnormally small.  For example,  a solar-mass BH
would have $\;\epsilon \sim 10^{-38}\;$ and we expect  that  $\; g_s^2 \lesssim \frac{1}{10}\;$, implying
that  $\;\varepsilon^2 <  10^{-38}\;$ (typically) and $\;\gamma^2 \lesssim \frac{1}{10}\;$.

It should then  be emphasized that the defrosted star background  must be viewed as a geometry which enables the frozen star to mimic some quantum effects of the polymer model by allowing it to be perturbed away from its bald equilibrium state. Clearly, any star for which $\;|g_{tt}|_{r=R}\; \sim \frac{1}{10}\;$ is not sufficiently compact to mimic a BH. The relevant physical quantities are the perturbations, which we discuss in the following sections, while the deformed background geometry is just an enabling agent.

It is also of interest to consider differences in the internal tortoise coordinate, $\;r_*=\int \frac{dr}{\sqrt{-g_{tt}g^{rr}}}\;$, which has important implications for the causal structure of the interior spacetime. In the case of the frozen star, this is  $\;r_*=\frac{r}{\varepsilon^2}\;$.
As for the defrosted star,  $\;r_*=\frac{R}{\gamma}\ln{\frac{r}{R}}\;$, which differs greatly from the linear relation of the frozen star model, being more
reminiscent of the familiar relation for the Schwarzschild geometry. In the analysis of Section~4, we find that redefining the tortoise coordinate as $\;r_*=R\ln{\frac{r}{R}}\;$, up to a suitable integration  constant, is a more convenient choice.

\section{Metric perturbations}

First note that, for the rest of the paper, including the Appendix,
we employ  units for which $\; 8 \pi G =1\;$.  The sole exception is Section~4.3,
where Newton's constant is briefly restored for clarity.

The metric perturbation $h_{\mu\nu}$ can be expanded in terms of
tensorial spherical harmonics of either even or  odd
parity. The Regge--Wheeler gauge \citep{regge1957stability} can be used
to simplify these expansions significantly. Our focus here will be on the even-parity  metric perturbations, as these are the ones that can couple to the scalar density and pressure perturbations \cite{thorne1967non} and so  the most relevant to an analysis that extends beyond the Cowling approximation.
The total perturbed metric in the Regge--Wheeler gauge is, to first order \citep{
thorne1967non,detweiler1985nonradial},
\begin{equation}
\tiny
g_{\mu\nu}^{\left(\text{even}\right)}\;=\;\left(\begin{array}{cccc}
-e^{2\Phi}\left(1+H_{0}Y_{\ell m}e^{i\omega t}\right) & -i\omega H_{1}Y_{\ell m}e^{i\omega t} & 0 & 0\\
-i\omega H_{1}Y_{\ell m}e^{i\omega t} & e^{2\Lambda}\left(1-H_{0}Y_{\ell m}e^{i\omega t}\right) & 0 & 0\\
0 & 0 & r^{2}\left(1-KY_{\ell m}e^{i\omega t}\right) & 0\\
0 & 0 & 0 & r^{2}\sin^{2}\theta\left(1-KY_{\ell m}e^{i\omega t}\right) \ \ \ \
\end{array}\right)\;.
\normalsize
\label{evenmetric}
\end{equation}

\normalsize
It is useful to introduce a fluid displacement vector of the form
$\;\xi^{i} = \frac{\delta u^{i}}{u^{t}}\;$, where
$\;i=1,2,3\;$ or $\;i=r,\theta,\phi\;$. Their Regge--Wheeler counterparts
are presented below in Eq.~(\ref{eq:xievendef}).

The four  radial functions characterizing the even-parity  metric perturbations, $H_{0},H_{1},H_{2},K$, and the pair of functions describing the  even-parity part of the displacement vector, $W, V$,
remain to be determined by the linearized equations of motion. These
emerge out of
the variation of the Einstein equations,
\begin{equation}
\delta G_{\mu\nu}\;=\;\delta T_{\mu\nu}\;,
\label{eq:pertEinstein}
\end{equation}
 along  with the variation of the conservation equations for the EM tensor,
\begin{equation}
\delta\left(\nabla_{\nu}T_{\phantom{\mu}\mu}^{\nu}\right)\;=\;0\;.
\label{eq:pertEnergyCons}
\end{equation}

As mentioned, the even-parity perturbations of the metric can be found in  Eq.~(\ref{evenmetric}),
whereas the even-parity fluid displacements are expressible as \cite{thorne1967non,detweiler1985nonradial}
\bea
\xi^{r}&=&\frac{e^{-\Lambda}W(r)}{r^{2}}e^{i\omega t}Y_{\ell m} \left(\theta,\phi\right),
\cr
\xi^{\theta}\;&=&\;-\frac{V(r)}{r^{2}}e^{i\omega t}\partial_{\theta}Y_{\ell m} \left(\theta,\phi\right),
\cr
\xi^{\phi}&=&-\frac{V(r)}{r^{2}\sin^{2}\theta} e^{i\omega t}\partial_{\phi}Y_{\ell m} \left(\theta,\phi\right)\;.
\label{eq:xievendef}
\eea

To relate spacetime and matter perturbations, one makes use of
Eqs.~(\ref{eq:pertEinstein}) and~(\ref{eq:pertEnergyCons}) such that
\begin{align}
T_{\;\;\mu}^{\nu} & \;=\;\text{diag}\left\{ -\rho,p,q,q\right\}\;.
\end{align}

For instance, Eq.~(\ref{eq:pertEinstein}) leads to a set of four initial-value equations
(see Section~\Subsecref{Perturbed-Einstein-equations}
in the Appendix for details),
\begin{align}
& H_{0}^{\prime}  +r^{-1}e^{2\Lambda}\left[1-r^{2}\rho+ \frac{\ell\left(\ell+1\right)}{2}+\frac{\sigma}{2}r^{2}\right]H_{0}\nonumber \\
 & \;=\;rK^{\prime\prime}+e^{2\Lambda}\left(3- \frac{5m\left(r\right)}{8\pi r}-\frac{r^{2}\rho}{2}\right)K^{\prime}+r^{-1}e^{2\Lambda}\left[1-\frac{\ell\left(\ell+1\right)}{2}-r^{2}\left(\rho+q\right)\right]K\nonumber \\
 & -r^{-1}\left[\frac{2\sigma}{r}- \frac{d\rho}{dr}\right] e^{\Lambda}W+r^{-1}e^{2\Lambda} \left(\rho+q\right)\ell\left(\ell+1\right)V+r^{-1}\left(\rho+p\right)e^{\Lambda}W^{\prime}\;,
 \label{eq:H0'}
\end{align}
\begin{align}
&\ell\left(\ell+1\right)H_{1}  \;=\;-2r\left[H_{0}+\left(r\Phi^{\prime}-1\right)K-rK^{\prime}\right]+2\left(\rho+p\right)e^{\Lambda}W\;,\label{eq:H1}
\end{align}
\begin{align}
\omega^{2}H_{1} & \;=\;-2e^{\Lambda+2\Phi}\left(p-q\right)\widehat{\delta k_{\theta}}-e^{2\Phi}\left(2\Phi^{\prime}H_{0}+H_{0}^{\prime}-K^{\prime}\right)\;,\label{eq:H1t}
\end{align}
\begin{equation}
H_{0}\;=\;H_{2}\;. \label{eq:H0H2}
\end{equation}
Here and in what follows, $\widehat{\delta k_{\theta}}$ denotes $\delta k_{\theta}$  stripped of its angular dependence,
\be
\;\delta k_{\theta}\left(r,\theta,\phi\right)=\widehat{\delta k_{\theta}}\left(r\right) \partial_{\theta} Y_{\ell m}\left(\theta,\phi\right)\;.
\label{stripped}
\ee
Similarly for other hatted quantities,
although in some cases it would be the spherical harmonic function that gets stripped off
rather than one that is differentiated.

\Eqref[s]{H1t} and~(\ref{eq:H0H2}) determine $\widehat{\delta k_{\theta}}$
and $H_{2}$, respectively, allowing for their elimination from all
the other equations (the latter, rather trivially).
The remaining two initial-value equations could, in principle, be used in a similar fashion. For instance,
Eq.~(\ref{eq:H1}) could   be used to eliminate
$H_{1}$ from the rest of the equations if supplemented by the relation
(see Eq.~(\ref{eq:H1'appendix}))
\begin{equation}
H_{1}^{\prime}\;=\;-r^{-1}e^{2\Lambda}\left(\frac{1}{2}r^{2}\left(p-\rho\right)+\frac{m\left(r\right)}{4\pi r}\right)H_{1}+e^{2\Lambda}\left(H_{0}+K-2\left(\rho+q\right)V\right)\;.\label{eq:H1'}
\end{equation}
Similarly, Eq.~(\ref{eq:H0'}) could be used to eliminate any one of $K,W,V$.
In practice, we will include this pair of initial-value relations  in our system of equations to be solved, leaving us with two extra unknowns.

Meanwhile,  the $_{\phantom{r}r}^{r}$ component of Eq.~(\ref{eq:pertEinstein})
gives us  the first of three propagation equations for what could have  been  the remaining three unknowns
($H_0$ and the ``other  two''  from the set  $K,W,V$),
\begin{align}
& e^{-2\Phi}\omega^{2}K  -e^{-2\Lambda}K^{\prime\prime}-2r^{-1} \left(e^{-2\Lambda}-\frac{r^{2}\left(\rho+p\right)}{4}\right) K^{\prime}\nonumber \\
 & -\left[r^{2}\left(\rho+p\right)-\ell\left(\ell+1\right)- \frac{\sigma}{2}r^{2}\left(1+\frac{dp}{d\rho}\right)\right] \frac{H_{0}}{r^{2}}+r^{2}\left(\rho+q\right) \left(1+\frac{dp}{d\rho}\right)\frac{K}{r^{2}}\nonumber \\
 & +\left(1+\frac{dp}{d\rho}\right) \left[\frac{2\sigma}{r}-\frac{d\rho}{dr}\right] \frac{e^{-\Lambda}W}{r^{2}}-\left(1+\frac{dp}{d\rho}\right)\left(\rho+p\right) \frac{e^{-\Lambda}W^{\prime}}{r^{2}}\nonumber \\
 & -\left(1+\frac{dp}{d\rho}\right) \left(\rho+q\right)\ell\left(\ell+1\right) \frac{V}{r^{2}}-2\omega^{2}e^{-2\left(\Lambda+\Phi\right)}\frac{H_{1}}{r}\nonumber \\
 & \;=\;0\;.\label{eq:Ktt}
\end{align}

 The variation of the energy-conservation equations, as given by
\begin{align}
0\;=\;\delta\left(T_{\mu\phantom{;\nu};\nu}^{\phantom{\nu}\nu}\right)
&= \partial_{t}\delta\left(\rho+q\right) u_{\mu}u^{t}+\partial_{\nu} \left(\rho+q\right)u_{\mu}\delta u^{\nu}\nonumber \\
 &
 +\delta\left(\rho+q\right)a_{\mu}+ \left(\rho+q\right) \left(u^{t}\delta\nabla_{t}u_{\mu}+\nabla_{\nu}u_{\mu}\delta u^{\nu}\right)
 \nonumber \\ &
 +\partial_{r}\delta\sigma k_{\mu}k^{r}+\partial_{\nu}\sigma\left(\delta k_{\mu}k^{r}\delta_{r}^{\nu}+k_{\mu}\delta k^{\nu}\right)\nonumber \\
 &+
 \delta\sigma\left(k^{r}\delta_{r}^{\nu}\nabla_{\nu}k_{\mu}+k_{\mu}\nabla_{\nu}k^{\nu}\right) +\partial_{\mu}\delta q\nonumber
 \\ &
 +\sigma\left(k^{r}\delta\left(\nabla_{r}k_{\mu}\right)+\delta k_{\mu}\left(\nabla_{\nu}k^{\nu}\right)+\nabla_{\nu}k_{\mu}\delta k^{\nu}+k_{\mu}\delta\left(\nabla_{\nu}k^{\nu}\right)\right)\;,
\end{align}
accounts for  the remaining two propagation equations (again, see the Appendix for details).

The first of these corresponds to the choice
$\;\mu=r\;$,
\begin{align}
0\;=\;\delta\left(T_{r\phantom{;\nu};\nu}^{\phantom{\nu}\nu}\right) & \;=\;\left\{ \partial_{r}+\Phi^{\prime}\left(\frac{\partial\rho}{\partial p}+1\right)+\frac{2}{r}\left(1-\frac{\partial q}{\partial p}\right)\right\} \delta p\nonumber \\
 & +\omega^{2}e^{-2\Phi}\left(\rho+p\right)H_{1}e^{i\omega t}Y_{\ell m}-\omega^{2}e^{-2\Phi}\left(\rho+p\right)r^{-2}e^{\Lambda}We^{i\omega t}Y_{\ell m}\nonumber \\
 & +\frac{1}{2}\left(\rho+p\right)H_{0}^{\prime}e^{i\omega t}Y_{\ell m}-\sigma K^{\prime}e^{i\omega t}Y_{\ell m}-\frac{\ell\left(\ell+1\right)}{r^{2}}\sigma e^{\Lambda}\widehat{\delta k}_{\theta}e^{i\omega t}Y_{\ell m}\nonumber \\
 & \;=\;\left\{ \partial_{r}+\Phi^{\prime}\left(\frac{\partial\rho}{\partial p}+1\right)+\frac{2}{r}\left(1-\frac{\partial q}{\partial p}\right)\right\} \left[-\frac{dp}{d\rho}\left(\rho+p\right)\frac{e^{-\Lambda}W^{\prime}}{r^{2}}\right.\nonumber \\
 & -\frac{dp}{d\rho}\left(\rho+q\right)\frac{\ell\left(\ell+1\right)}{r^{2}}V+\frac{dp}{d\rho}\frac{2\sigma}{r^{3}}e^{-\Lambda}W\nonumber \\
 & \left.-\frac{dp}{dr}e^{-\Lambda}\frac{W}{r^{2}}+\frac{1}{2}\frac{dp}{d\rho}\sigma H_{0}+\frac{dp}{d\rho}\left(\rho+q\right)K\right]\nonumber \\
 & +\omega^{2}e^{-2\Phi}\left(\rho+p\right)H_{1}-\omega^{2}e^{-2\Phi}\left(\rho+p\right)r^{-2}e^{\Lambda}W\nonumber \\
 & +\frac{1}{2}\left(\rho+p\right)H_{0}^{\prime}-\sigma K^{\prime}-\ell\left(\ell+1\right)\sigma e^{\Lambda}r^{-2}\widehat{\delta k}_{\theta}\;,
 \label{eq:Wtt}
\end{align}
and the second to $\;\mu=\theta\;$,
\begin{align}
0\;=\;\delta\left(T_{\theta\phantom{;\nu};\nu}^{\phantom{\nu}\nu}\right) & \;=\;\partial_{\theta}\delta q+\left(\rho+q\right)e^{-2\Phi}\omega^{2}Ve^{i\omega t}\partial_{\theta}Y_{\ell m}+\frac{1}{2}\left(\rho+p\right)H_{0}e^{i\omega t}\partial_{\theta}Y_{\ell m}\nonumber \\
 & +e^{-\Lambda}\sigma\partial_{r}\delta k_{\theta}+e^{-\Lambda}\sigma\delta k_{\theta}\left(\frac{2}{r}+\Phi^{\prime}+\partial_{r}\ln\sigma\right)\nonumber \\
 & \;=\;-\frac{dq}{d\rho}\left[\left(\rho+p\right)\frac{e^{-\Lambda}W^{\prime}}{r^{2}}+\left(\rho+q\right)\frac{\ell\left(\ell+1\right)}{r^{2}}V\right]\nonumber \\
 & +\frac{dq}{d\rho}\left(\frac{2\sigma}{r}-\frac{d\rho}{dr}\right)e^{-\Lambda}\frac{W}{r^{2}}+\frac{dq}{d\rho}\left(\rho+q\right)K\nonumber \\
 & +\left(\rho+q\right)e^{-2\Phi}\omega^{2}V+\frac{H_{0}}{2}\left[\rho+p+\sigma\frac{dq}{d\rho}-e^{-2\Lambda}\left(2\Phi^{\prime}\left(\Lambda^{\prime}-\frac{\partial_{r}\sigma}{\sigma}\right)-2\Phi^{\prime\prime}\right)\right]\nonumber \\
 & +\frac{e^{-2\Lambda-2\Phi}}{2}\left\{ e^{2\Phi}H_{0}^{\prime\prime}-e^{2\Phi}K^{\prime\prime}-\left(\frac{\partial_{r}\sigma}{\sigma}+\Lambda^{\prime}+2\Phi^{\prime}\right)\omega^{2}H_{1}+\omega^{2}H_{1}^{\prime}\right.\nonumber \\
 & \left.-\left[\Lambda^{\prime}-2\Phi^{\prime}+\frac{\partial_{r}\sigma}{\sigma}\right]e^{2\Phi}H_{0}^{\prime}+\left(\Lambda^{\prime}+\frac{\partial_{r}\sigma}{\sigma}\right)e^{2\Phi}K^{\prime}\right\} \nonumber \\
 & +\frac{e^{-2\Lambda-2\Phi}}{2}\left[\omega^{2}H_{1}+e^{2\Phi}\left(2\Phi^{\prime}H_{0}+H_{0}^{\prime}-K^{\prime}\right)\right]\left(\frac{2}{r}+\Phi^{\prime}+\partial_{r}\ln\sigma\right).\label{eq:Vtt}
\end{align}
Note that these last two equations have  introduced matter fluctuations into the mix;
namely, $\delta \rho$, $\delta q$ and $\delta p$. Importantly, $\delta\rho$ and $\delta q$
can be expressed in terms of $\delta p$, as we do
in the next four equations below,  given that the
equation-of-state relations
are known.

It is convenient for our upcoming  analysis to rewrite some  of the above
equations. Using  Eq.~(\ref{eq:H0'withdeltarho}) for $\delta \rho$ along with
$\;\delta\rho=\left(\partial\rho/\partial p\right)\delta p\;$, we
obtain  a revised form for Eq.~(\ref{eq:H0'}),
\begin{align}
2r^{2}\frac{\partial\rho}{\partial p}\delta p & \;=\;e^{i\omega t}Y_{\ell m}\left\{ \left(2-\ell\left(\ell+1\right)\right)K+ \left[2e^{-2\Lambda}\left(2r\Lambda^{\prime}-1\right)-\ell\left(\ell+1\right)\right]H_{0}\right. \nonumber \\
 & \left.-2e^{-2\Lambda}r\left(H_{0}^{\prime}+\left(r\Lambda^{\prime} -3\right)K^{\prime}-rK^{\prime\prime}\right)\right\} \;.
 \label{eq:eqn1}
\end{align}

Also,  Eq.~(\ref{eq:K''withdeltap}) for $\delta p$ allows us to reexpress
Eq.~(\ref{eq:Ktt}) as
\begin{align}
2r^{2}\delta p\;= & \;e^{i\omega t}Y_{\ell m} \left\{ -\left(2-\ell\left(\ell+1\right)+2r^{2}e^{-2\Phi}\omega^{2}\right)K +4r\omega^{2}e^{-2\left(\Lambda+\Phi\right)}H_{1}\right.\nonumber \\
 &
 \left.+2e^{-2\Lambda} \left[rH_{0}^{\prime}-r\left(1+r\Phi^{\prime}\right)K^{\prime}\right] +\left[2e^{-2\Lambda}\left(1+2r\Phi^{\prime}\right) -\ell\left(\ell+1\right)\right]H_{0}\right\}\;.\label{eq:eqn2}
\end{align}

Furthermore, Eqs.~(\ref{eq:Wtt}) and~(\ref{eq:Vtt}) can  be
rewritten by  using Eq.~(\ref{eq:H1t})  to eliminate $\delta k_{\theta}$. Respectively,
\begin{align}
&\left\{ \partial_{r}+\Phi^{\prime} \left(\frac{\partial\rho}{\partial p}+1\right) +\frac{2}{r}\left(1-\frac{\partial q}{\partial p}\right)\right\} \widehat{\delta p}
 \;=\; -e^{-2\Phi}\omega^{2}H_{1}\left[\frac{\ell\left(\ell+1\right)}{2r^{2}}+\left(\rho+p\right)\right] \nonumber \\
 & +\omega^{2}e^{-2\Phi}\left(\rho+p\right)\frac{e^{\Lambda}W}{r^{2}} -\frac{1}{2}\left[\frac{\ell\left(\ell+1\right)}{r^{2}}+\left(\rho+p\right)\right]H_{0}^{\prime} \nonumber \\
 & +\left[\sigma+\frac{\ell\left(\ell+1\right)}{2r^{2}}\right]K^{\prime} -\frac{\ell\left(\ell+1\right)}{r^{2}}\Phi^{\prime}H_{0}\;
 \label{eq:eqn3}
\end{align}
and
\begin{align}
&\frac{\partial q}{\partial p}\widehat{\delta p} \;=\;-\left(\rho+q\right)e^{-2\Phi}\omega^{2}V-\frac{1}{2}\left(\rho+p\right)H_{0}\nonumber \\
 & -\frac{1}{2}\omega^{2}\left(\Lambda^{\prime}+\Phi^{\prime}-\frac{2}{r}\right)e^{-2\Lambda -2\Phi}H_{1}+\frac{1}{2}\omega^{2}e^{-2\Lambda-2\Phi}H_{1}^{\prime} \nonumber \\
 & +e^{-2\Lambda}\left(\Phi^{\prime\prime}H_{0} +\Phi^{\prime}H_{0}^{\prime}+\frac{1}{2}H_{0}^{\prime\prime} -\frac{1}{2}K^{\prime\prime}\right)\nonumber \\
 & +e^{-2\Lambda}\left(\Phi^{\prime}H_{0}+ \frac{1}{2}H_{0}^{\prime} -\frac{1}{2}K^{\prime}\right)\left(\frac{2}{r}+\Phi^{\prime}-\Lambda^{\prime}\right)\;.
 \label{eq:eqn4}
\end{align}


Equations~(\ref{eq:eqn1})--(\ref{eq:eqn4}),
along with Eqs.~(\ref{eq:H1}) and~(\ref{eq:H1'}),
as well as the following expression for $\delta p$ (see  Eq.~(\ref{eq:deltap})),
\begin{align}
\widehat{\delta p} & \;=\;\frac{dp}{d\rho}\left[\frac{2\sigma}{r}-\frac{d\rho}{dr}\right]e^{-\Lambda}\frac{W}{r^{2}} -\frac{dp}{d\rho}\left(\rho+p\right)\frac{e^{-\Lambda}W^{\prime}}{r^{2} }-\frac{dp}{d\rho}\left(\rho+q\right)\frac{\ell\left(\ell+1\right)}{r^{2}}V
\nonumber \\
 & +\frac{1}{2}\frac{dp}{d\rho}\sigma H_{0}+\frac{dp}{d\rho}\left(\rho+q\right)K\;,
 \label{eq:eqn5}
\end{align}
form a set of seven equations for  the seven variables $H_{0},H_{1},K,V,W,\delta p$ and the frequency $\omega$. This will be the foundation for the upcoming analysis on
the oscillation modes of the defrosted star.

\section{Defrosting a star}

We will now proceed to apply the  formalism of Section~3 to our defrosted star model. We calculate the equations and the solutions to leading order in $\gamma$.

\subsection{Perturbation equations for the defrosted star}

Let us recall the set of seven equations from the end of Section~3; namely,
Eqs. (\ref{eq:eqn1}-\ref{eq:eqn4}), (\ref{eq:H1}), (\ref{eq:H1'}) and (\ref{eq:eqn5}).
Plugging in the expressions for $\Lambda,\Phi$ and $\rho,p,q$ for the defrosted star model (see Section~2.2), we obtain the following set of seven respective relations
:

\begin{subequations}
\begin{equation}
-2r^{2}\widehat{\delta p}\;=\;\left(2-\ell\left(\ell+1\right)\right)K-\ell\left(\ell+1\right)H_{0}\;,\label{eq:KH01}
\end{equation}
\begin{align}
-2r^{2}\widehat{\delta p}\;=\; & \left(2-\ell\left(\ell+1\right)\right)K+\ell\left(\ell+1\right)H_{0}\;,\label{eq:KH02}
\end{align}

\begin{align}
\partial_{r}\left(r^{2}\widehat{\delta p}\right)  \;=\;-\gamma\widetilde{\omega}^{2} & \frac{\ell\left(\ell+1\right)}{2r^{2}}H_{1} -\frac{\ell\left(\ell+1\right)}{2}\frac{\partial_{r}\left(r^{2}H_{0}\right)}{r^{2}}- \frac{1}{2}\left(2-\ell\left(\ell+1\right)\right)K^{\prime}\;,
\end{align}
\begin{align}
-r^{2}\widehat{\delta p} & \;=\;-\widetilde{\omega}^{2} \frac{V}{r^{2}}+H_{0}+\frac{5}{2}rH_{0}^{\prime}+\frac{1}{2}r^{2}H_{0}^{\prime\prime} \nonumber \\
 & +\frac{1}{2}\gamma\widetilde{\omega}^{2} \frac{\partial_{r}\left(rH_{1}\right)}{r}-\frac{3}{2}rK^{\prime} -\frac{1}{2}r^{2}K^{\prime\prime}\;,
 \label{eq:dpVKK'K''}
\end{align}
\begin{equation}
\ell\left(\ell+1\right)H_{1}\;=\;-2rH_{0}+ 2r^{2}K^{\prime}+\frac{4\gamma^{1/2}}{r^{2}}W\;,
\label{eq:H1H0K'W}
\end{equation}
\begin{equation}
\gamma\frac{\partial_{r}\left(rH_{1}\right)}{r}\;=\;H_{0}+K-\frac{2V}{r^{2}}\;,
\label{eq:H1H0KV}
\end{equation}
\begin{align}
r^{2}\widehat{\delta p} & \;=\;\ell\left(\ell+1\right)\frac{V}{r^{2}}+\frac{1}{2}H_{0}-K\;,
\label{eq:dpVH0K}
\end{align}
\end{subequations}
where
\be
\widetilde{\omega}^{2}\;=\; \frac{R^{2}\omega^{2}}{\gamma^{2}}\;.
\label{eq:wtilde}
\ee

The combination of Eqs.~(\ref{eq:KH01}) and~(\ref{eq:KH02}) inevitably
leads to
\begin{equation}
2\ell\left(\ell+1\right)H_{0}\;=\;0\;,
\end{equation}
which means that, for any $\;\ell>0\;$, $\;H_{0}=0\;$ to leading order in $\gamma$.

Restricting to $\;\ell\geq 2\;$  in Eq.~(\ref{eq:H1H0K'W})
informs us  that $\;H_{1}\sim\mathcal{O}\left(\gamma\right)\;$ and, therefore, the terms
containing $\gamma H_{1}$ are negligible in the rest of the equations.

In light of our new  knowledge, Eq.~(\ref{eq:dpVKK'K''})  reduces to
\begin{equation}
-r^{2}\widehat{\delta p}\;=\;-\widetilde{\omega}^{2}\frac{V}{r^{2}} -\frac{3}{2}rK^{\prime}-\frac{1}{2}r^{2}K^{\prime\prime}\;,
\label{eq:preEulerK}
\end{equation}
where $K$ and $V$ are directly related through Eq.~(\ref{eq:H1H0KV}),
\begin{equation}
\frac{V}{r^{2}}\;=\;\frac{K}{2}\;,
\label{eq:VtoK}
\end{equation}
and where, by virtue of Eq.~(\ref{eq:KH02}), $K$ and $r^{2}\widehat{\delta p}$
are also related through
\begin{equation}
r^{2}\widehat{\delta p}\;=\;\frac{1}{2}\left(\ell\left(\ell+1\right)-2\right)K\;.
\end{equation}

The previous relation allows us to recast  Eq.~(\ref{eq:preEulerK}) into a simple equation for $K$,
\begin{equation}
r^{2}K^{\prime\prime}+3rK^{\prime}+ \left[\widetilde{\omega}^{2}-\left(\ell\left(\ell+1\right)-2\right)\right]K\;=\;0\;.
\label{eq:EulerK}
\end{equation}

We will discuss the solutions of Eq.~(\ref{eq:EulerK}) in Section~4.3.

\subsection{Back to the Cowling approximation}

One might expect that setting all metric perturbations to zero in
Eqs. (\ref{eq:H0'}-\ref{eq:Vtt}) would lead to the same results
as those obtained
under  the Cowling approximation, where the working assumption is that
all gravitational perturbations are negligibly small and so effectively
vanishing.
However, when $H_{0},H_{1}$ and $K$ vanish, Eq.~(\ref{eq:H1})
then implies that $W$ vanishes. This dubious result, along with Eq.~(\ref{eq:H0'}),
would further imply that  $V$ also vanishes. But, as firmly established in
other works such as \cite{fluctuations}, this is not what one finds when the Cowling approximation is correctly applied.

The resolution to this nonsensical result is the observation that
the initial-value and propagation equations, as presented in
the previous section, are describing the \textbf{back-reaction}
on the metric   perturbations and  not their source.
Indeed, it was argued
in \citep{ipser1992pulsations} that the Cowling approximation is only
valid when the self-gravitation of the modes is negligible. This
is not necessarily the case in the current situation, which means
that one cannot connect the current framework to the Cowling approximation
simply by setting $H_{0},H_{1}$ and $K$ to zero.

\subsection{Spectrum of non-radial oscillations of the defrosted star}

Let us next determine the solutions of Eq.~(\ref{eq:EulerK})
by imposing suitable boundary conditions.

To help facilitate this process, it is useful to introduce a new radial coordinate in the star's interior, $\;r \to r_*^{in}\;$, which will serve as   the analogue of the tortoise coordinate  in the Schwarzschild exterior, $\;r_*^{out}= r+ 2MG\ln\left(1-\frac{2MG}{r}\right)\;$.  As per our definition
for the interior tortoise coordinate
at the end of Section~2.3, $\;r_{\ast}^{in}=\ln{\frac{r}{R}}+{\rm Const.}\;$, the line element for the defrosted star,
\begin{equation}
ds^{2}_{DS}\;=\;-\gamma\left(\frac{r}{R}\right)^{2}dt^{2}+\frac{1}{\gamma}dr^{2}+r^{2}d\Omega^{2}\;,
\end{equation}
transforms into
\begin{equation}
ds^{2}_{DS}\;=\;\left(\frac{r}{R}\right)^{2}\left[-\gamma dt^{2}+\frac{1}{\gamma}\left(dr_*^{in}\right)^{2}+r^{2}d\Omega^{2}\right]\;.
\label{eq:r*in}
\end{equation}

This definition for the internal tortoise coordinate  is motivated by the expectation that the modes are non-relativistic as suggested by  Eq.~(\ref{eq:wtilde}). Additionally, we require that $r_*^{in}$ joins $r_*^{out}$ in a continuous way at $\;r=R\simeq 2MG (1+\gamma)\;$, where $\;r_*^{out}\simeq  R + 2MG\ln{\gamma}\;$ .  This leads to
\be
r_*^{in}\;=\; R \ln \left(\frac{r}{R}\right)+ R\left(1+\ln\gamma\right)\;.
\label{r*in2}
\ee

In these coordinates, Eq.~(\ref{eq:EulerK}) becomes
\be
\partial_*^2 K + \frac{2}{R} \partial_* K +\frac{1}{R^2}\left[\widetilde{\omega}^{2}-\left(\ell\left(\ell+1\right)-2\right)\right] K \;=\;0\;,
\label{eom*K}
\ee
for which the general solution takes the form
\begin{equation}
K_{in}\left(r_*^{in}\right)\;=\; \frac{1}{r}\left( A~ e^{\hbox{$ i k_{in} r_*^{in}$}}
+ B~ e^{\hbox{$- i k_{in} r_*^{in}$}} \right)\;,
\end{equation}
where $A$ and $B$ are complex constants and the frequency is given by
\begin{equation}
\omega_{in}^{2} \;=\;\gamma^2 \left[k_{in}^2 +\frac{1}{R^2}\left(\ell(\ell+1)-2\right)\right]\;.
\label{omekay}
\end{equation}

The boundary conditions that are imposed on $K_{in}$ determine the coefficients $A$ and $B$.
First, we impose the standard boundary condition at $\;r \to 0\;$, which is the vanishing of the current at the star's center, which means that the ingoing and outgoing parts of the solution are equal and opposite, leading to  $\;B=-A\;$ and thus
\begin{equation}
K_{in} \;=\;A~\frac{\sin\left(k_{in} \,r_*^{in}\right)}{r}\;.
\end{equation}

At the $\;r=R\;$ outer boundary, we impose the standard condition of an outgoing wave, along with the continuity of the solution and its first-order derivative across the surface.
The external solution then takes on its usual  form of  $\;K_{out}(r_*^{out})\sim\frac{1}{r}
e^{- i(\omega_{out}t - k_{out} r_*^{out})}\;$ with $\;k_{out}=\omega_{out}\;$.
First, comparing the time dependence of the inner and outgoing solutions,
we observe that
\begin{equation}
\omega_{in}\;=\;\omega_{out}\;.
\end{equation}

Next, continuity of the solutions and of their logarithmic $r_*$-derivative at
the surface leads to the following relation:
\begin{equation}
\cot\left(k_{in} R\left(1+\ln\gamma \right)\right)\;=\; i \frac{k_{out}}{k_{in}} = i \frac{\omega_{out}}{k_{in}}=i\frac{\omega_{in}}{k_{in}} \;,
\label{kay}
\end{equation}
which fixes the allowed values of $k_{in}$ and thus
those of $\omega_{in}$.
This form of  relation  was anticipated in \cite{collision} in the context of  the closely related polymer model.

Equation~(\ref{kay}) implies that the real part of $k_{in}$ satisfies
\be
 \text{Re}\left(k_{in}\right)\;= \;m \frac{\pi}{2R \left(1+\ln\gamma\right)}\;,\;\;\; m \;=\;\pm 1,\ 3,\ 5,\ \dots\;,
\ee
 so that the  real part of $\omega_{in}$ is given by (also see Eq.~(\ref{omekay}))
\be
\left[\text{Re}\left(\omega_{in}\right)\right]^2 \;=\;\frac{\gamma^2}{R^2} \left[ \frac{m^2 \pi^2}{4 \left(1+\ln\gamma\right)^2} +\ell(\ell+1)-2\right]\;,\;\;\; m = 1,\ 3,\ 5,\ \dots\;.
\ee
The result is therefore that the sound velocity is non-relativistic and scales as $\gamma$, $\;v_{sound}=  \text{Re}\left(\omega_{in}\right)/\text{Re}\left(k_{in}\right)\sim \gamma\;$.

Meanwhile, the imaginary part of the frequency goes as
\be
 \text{Im}\left(\;\omega_{in}\right)\;=\; \frac{1}{2 R(1+\ln\gamma)} \ln\left[\frac{\frac{k_{in}}{\omega_{in}}+1}{\frac{k_{in}}{\omega_{in}}-1}\right]\;.
\ee

It has been assumed throughout the discussion that $\gamma$ is a small parameter; hence, we may expand the imaginary part of the frequency in
$\;\frac{\omega_{in}}{k_{in}}\sim \gamma\;$.
The result is
\be
 \text{Im}\left(\;\omega_{in}\right)\;=\; \frac{1}{ R(1+\ln\gamma)} \left(\frac{\text{Re}\left(\omega_{in}\right)}{\text{Re}\left(k_{in}\right)}\right)^2.
\label{spectra2a}
\ee
The interesting feature here is that the imaginary part is parametrically
smaller than the real part, since the former scales as $\gamma^2$ while the the latter scales as $\gamma$.  We arrive at the conclusion that  non-relativistic modes do couple to waves in the  external spacetime and should leak out of the interior at a slow rate which goes as $\;\text{Im}\left(\omega\right) \sim \gamma^2\;$. This is exactly what was found in \cite{collision} for the polymer model; except that, there, $ g_s^2\;$ is the perturbative parameter rather than $\gamma^2$.

In a similar discussion in \cite{collision}, $\dfrac{k_{in}}{\omega_{in}}$ was denoted by $n$ (the refraction index). In that discussion, two limiting cases were considered: $\;n\gtrsim 1\;$  and $\;n\gg 1\;$, which
now translate
into
$\;\dfrac{1}{\gamma}\gtrsim 1\;$ and  $\; \dfrac{1}{\gamma}\gg 1\;$. Momentarily, to simplify the notation, the $in$ subscripts will be implied and $\;\dfrac{k_{in}}{\omega_{in}}$ will be denoted by $n$.

Then, for the former, nearly relativistic case, one finds that
\be
\text{Im}\left(\omega\right)\;\simeq\; \frac{1}{2R(1+\ln\gamma)}\ln{\left(\frac{2}{n-1}\right)}\;.
\label{spectra2}
\ee
The logarithm in the imaginary part  appears to diverge in the relativistic limit,
$\;n \to 1\;$. However, as discussed in \cite{collision}, this is
just an apparent problem because the amplitude of these waves scales as
$\;A(r=R)\sim (n-1)^{t/2R}\;$, which goes rapidly to zero in time as $n$ approaches unity.
The suppression of relativistic fluid modes is actually  ubiquitous in the literature \cite{inversecowling,Kokk1,QNMBH,QNMBH2,Kokk2,gravi,Cardoso,Cardoso:2016oxy}.

For the latter case of  a large index, $\;n \gg 1\;$, one can expand the logarithm in terms of $\frac{1}{n}$,
which leads to
\be
\text{Im}\left(\omega\right)\;=\; \left[\frac{1}{R(1+\ln\gamma)\; n^2}
+ {\cal O}\left(\frac{1}{n^4}\right)\right]\;,
\label{spectra2b}
\ee
which is the result found in Eq.~(\ref{spectra2a}).

This feature was anticipated not only because of the analysis in \cite{collision} but  because of a physical argument in \cite{ridethewave}. The gist of the argument is the following: Because their frequency is parametrically small,   these modes are viewed by an external observer as having a parametrically long wavelength, $\;\lambda \sim n R\;$, when compared to the size $R$ of the compact object. This reduces the transmission cross-section through a surface of area $R^2$ by a factor of $\;\dfrac{R^2}{\lambda^2}\sim \dfrac{1}{n^2}  \;$. As the  cross-section determines the power loss, it follows that $\;\dfrac{dE}{dt}\sim\dfrac{R^2}{\lambda^2}\sim \dfrac{1}{n^2} \;$, which in turn determines the inverse of the damping-time scale as $\;\dfrac{1}{\tau} \sim \dfrac{1}{n^2}\;$. In other words,   $\; \text{Im}\left(\omega\right) \sim \dfrac{1}{n^2}\;$.

\section{Discussion}

We have calculated the  spectrum  for the even-parity non-radial oscillatory modes of  a defrosted star, which is our name for a suitably deformed ---  or excited --- version of a BH mimicker whose ground state is described by the  frozen star model. Due to the ultrastability of the frozen star geometry, a deformation of the background solution  is necessary for the star to support perturbative, oscillatory  modes.
As the star's  ultrastability is directly linked to its radial pressure being maximally negative, it is straightforward to quantify the deformation in  terms of the deviation from maximal negative pressure, as denoted by
$\gamma$.
What we have found is that the mode frequencies scale  with this small  strength-of-deformation parameter  $\gamma$, whereas their imaginary parts scale as $\gamma^2$. The lifetime of these modes is thus predicted to be parametrically long.
We expect these characteristics of the oscillation modes to be reflected in the properties of the spectrum of emitted GWs from excited frozen stars, which may  have important implications when it comes to the potential observability
of these emissions via GW detectors.

The current analysis follows an earlier one \cite{fluctuations} that aimed at similar goals but did so by  applying the Cowling approximation, for  which  the interior oscillatory modes are not coupled to the external geometry. This coupling is a necessary step for the calculation of the production  of GWs. Relaxing the approximation in the case of  highly anisotropic stars is a technical challenge in its own right. It is then our hope that the detailed presentation of the formalism, which applies to a static, spherically symmetric but otherwise-generic anisotropic star, will prove to have merit independently of any particular model.

The main results that we obtained here were forecast by an earlier study which determined the mode spectrum  for our  polymer model  \cite{collision}, as well as by a later article which provided  physical arguments in support of those earlier findings \cite{ridethewave}. As the polymer model is supposed to provide a microscopic description of the frozen star, this should not have been a surprise. On the other hand, the  spectral derivation in the polymer framework seemed  at times to be rather heuristic. The   current analysis is thus  serving to  vindicate the earlier one, as well as substantiating our contention that the polymer and frozen star models are really different descriptions of the same class of objects, and likewise for their excited states.

As most astrophysical BHs are rotating at large fractions of the speed of light,
it will be difficult to make a precise connection with the empirical data until rotation is  formally incorporated into the calculation. Nevertheless, we expect that the lifetime of the modes will retain its scaling, while the frequencies will be brought up to the rotation frequency of a BH.  Fortunately, a model that will allow us to verify this expectation for the rotating frozen star is already available \cite{NotSteveKerr}, and its defrosted counterpart should  not be too far behind. It will also be interesting to recast this problem in the formalism of a recently introduced open-string description of the frozen star geometry \cite{FluxUstat}, as this framework includes a matter Lagrangian of the Born--Infeld class.

\section*{Acknowledgments}
The research is supported by the German Research Foundation through a German-Israeli Project Cooperation (DIP) grant ``Holography and the Swampland'' and by VATAT (Israel planning and budgeting committee) grant for supporting theoretical high energy physics. The research of AJMM received support from an NRF Evaluation and Rating Grant 119411 and a Rhodes  Discretionary Grant SD07/2022. AJMM thanks Ben Gurion University for their hospitality during past visits.

\appendix
\section{Even-parity, linear perturbations}

Here, we are working to  first order in  the perturbations, a  point that will only sometimes be made explicit. The notation $\mathcal{O}(\delta^n)$ is to be understood as indicating the order of a given expression in terms of the relevant  perturbative parameter, {\em e.g.},  $H_0, H_1, K, W, V, \gamma$, {\em etc}.

\subsection{Perturbation to the fluid velocity $u^{\mu}$ }

The total perturbed metric in the Regge--Wheeler gauge is, to first order{\footnotesize{},
\begin{equation}
\tiny
g_{\mu\nu}^{\left(\text{even}\right)}\;=\;\left(\begin{array}{cccc}
-e^{2\Phi}\left(1+H_{0}Y_{\ell m}e^{i\omega t}\right) & -i\omega H_{1}Y_{\ell m}e^{i\omega t} & 0 & 0\\
-i\omega H_{1}Y_{\ell m}e^{i\omega t} & e^{2\Lambda}\left(1-H_{0}Y_{\ell m}e^{i\omega t}\right) & 0 & 0\\
0 & 0 & r^{2}\left(1-KY_{\ell m}e^{i\omega t}\right) & 0\\
0 & 0 & 0 & r^{2}\sin^{2}\theta\left(1-KY_{\ell m}e^{i\omega t}\right) \ \ \ \
\end{array}\right)\;,
\normalsize
\label{evenmetric1}
\end{equation}
\ \\
\begin{equation}
\tiny
g^{\mu\nu\left(\text{even}\right)}\;=\;\left(\begin{array}{cccc}
-e^{-2\Phi}\left(1-H_{0}Y_{\ell m}e^{i\omega t}\right) & -i\omega e^{-2\left(\Lambda+\Phi\right)}H_{1}Y_{\ell m}e^{i\omega t} & 0 & 0\\
-i\omega e^{-2\left(\Lambda+\Phi\right)}H_{1}Y_{\ell m}e^{i\omega t} & e^{-2\Lambda}\left(1+H_{0}Y_{\ell m}e^{i\omega t}\right) & 0 & 0\\
0 & 0 & r^{-2}\left(1+KY_{\ell m}e^{i\omega t}\right) & 0\\
0 & 0 & 0 & r^{-2}\sin^{-2}\theta\left(1+KY_{\ell m}e^{i\omega t}\right)
\end{array}\right)\;.
\normalsize
\label{evenmetricup}
\end{equation}
}

\normalsize
Choosing to work in the rest frame, one has
\begin{equation}
u^{\mu}\;=\;u^{t}\delta_{t}^{\mu}\;,\qquad u_{t}\;=\;g_{t\mu}u^{\mu}=g_{\mu t}u^{t}\delta_{t}^{\mu}\;=\;g_{tt}u^{t}\;,
\end{equation}
and thus  can obtain all non-vanishing components of the velocity,
\begin{align}
u_{t} & \;=\;g_{tt}u^{t}\;,\\
u_{r} & \;=\;g_{rt}u^{t}\;.
\end{align}

Using the normalization of the velocity, one can show that
\begin{equation}
-1\;=\;u_{\mu}u^{\mu}\;=\;g_{\mu\nu}\delta_{t}^{\nu}\delta_{t}^{\mu}\left(u^{t}\right)^{2}\;,
\end{equation}
which leads to
\begin{equation}
u^{t}\;=\;\sqrt{\frac{1}{g_{tt}}}\;=\;\sqrt{\frac{1}{e^{2\Phi}\left(1+H_{0}Y_{\ell m}e^{i\omega t}\right)}}\;\approx\; e^{-\Phi}\left(1-\frac{1}{2}H_{0}e^{i\omega t}Y_{\ell m}\right)\;,
\end{equation}
from which  $\;u_{\alpha}=g_{\alpha t}u^{t}\;$ can be used to yield  explicit expressions
for the remaining components of the velocity,
\begin{equation}
u_{t}\;\approx\;-e^{\Phi}\left(1+\frac{1}{2}H_{0}e^{i\omega t}Y_{\ell m}\right)\;,\qquad u_{r}\;\approx\;-i\omega e^{-\Phi}H_{1}Y_{\ell m}e^{i\omega t}\;,\qquad u_{\theta,\phi}\;=\;0\;.
\end{equation}

By the definition provided  in
Section~3 for the displacement vector $\xi^i$,
the radial and angular variations of the velocity are, to first order,
\begin{align}
\delta u^{r} & \;=\;u^{t}\partial_{t}\xi^{r}\nonumber \\
 & \;=\;i\omega r^{-2}e^{-\left(\Phi+\Lambda\right)}We^{i\omega t}Y_{\ell m}+\mathcal{O}\left(\delta^{2}\right)\;,
\end{align}
\begin{align}
\delta u^{\theta} & \;=\;u^{t}\partial_{t}\xi^{\theta}\nonumber \\
 & \;=\;-i\omega r^{-2}e^{-\Phi}Ve^{i\omega t}\partial_{\theta}Y_{\ell m}+\mathcal{O}\left(\delta^{2}\right)\;,
\end{align}
\begin{align}
\delta u^{\phi} & \;=\;u^{t}\partial_{t}\xi^{\phi}\nonumber \\
 & \;=\;-i\omega r^{-2}\sin^{-2}\theta e^{-\Phi}Ve^{i\omega t}\partial_{\phi}Y_{\ell m}+\mathcal{O}\left(\delta^{2}\right)\;.
\end{align}

Lowering the indices on the previous variations, one obtains
\begin{align}
\delta u_{r} &\; =\;\delta\left(u^{\mu}g_{\mu r}\right)\nonumber \\
 & \;=\;\delta u^{\mu}g_{\mu r}+u^{\mu}\delta g_{\mu r}\nonumber \\
 & \;=\;\delta u^{r}g_{rr}+u^{t}h_{tr}\nonumber \\
 & \;\approx\; i\omega r^{-2}e^{-\left(\Phi-\Lambda\right)}We^{i\omega t}Y_{\ell m}-i\omega e^{-\Phi}H_{1}Y_{\ell m}e^{i\omega t}\;,
\end{align}
\begin{align}
\delta u_{\theta} & \;=\;\delta\left(u^{\mu}g_{\mu\theta}\right)\nonumber \\
 & \;=\;\delta u^{\theta}g_{\theta\theta}+u^{\theta}\delta g_{\theta\theta}\nonumber \\
 & \;=\;-i\omega r^{-2}e^{-\Phi}Ve^{i\omega t}Y_{\ell m}r^{2}\left(1-Ke^{i\omega t}Y_{\ell m}\right)\nonumber \\
 & \;\approx\;-i\omega e^{-\Phi}Ve^{i\omega t}\partial_{\theta}Y_{\ell m}\;,
\end{align}
\begin{align}
\delta u_{\phi} & \;=\;\delta u^{\phi}g_{\phi\phi}\nonumber \\
 & \;\approx\;-i\omega e^{-\Phi}Ve^{i\omega t}\partial_{\phi}Y_{\ell m}\;.
\end{align}

\subsection{Perturbations to the radial vector $k^\mu$}

Using the normalization condition  $\;k^{\mu}k_{\mu}=1\;$ and that, in the rest
frame, $\;k^{\mu}=\delta_{r}^{\mu}k^{r}\;$ is a purely radial vector,
one  finds that
\begin{align}
k^{r}\;=\; & \sqrt{\frac{1}{g_{rr}}}\nonumber \\
 &\; =\;\sqrt{\frac{1}{e^{2\Lambda}\left[1-H_{0}e^{i\omega t}Y_{\ell m}\right]}}\nonumber \\
 &\; \approx\; e^{-\Lambda}\left[1+\frac{1}{2}H_{0}e^{i\omega t}Y_{\ell m}\right]\;,
\end{align}
as well as
\begin{align}
k_{r} & \;=\;g_{rr}k^{r}\nonumber \\
 & \;=\;\sqrt{e^{2\Lambda}\left[1-H_{0}e^{i\omega t}Y_{\ell m}\right]}\nonumber \\
 & \;\approx\; e^{\Lambda}\left[1-\frac{1}{2}H_{0}e^{i\omega t}Y_{\ell m}\right]\;, \\
k_{t} & \;=\;g_{rt}k^{r}\nonumber \\
 & \;\approx\;-i\omega e^{-\Lambda}H_{1}e^{i\omega t}Y_{\ell m}\sim\mathcal{O}\left(\delta\right)\;.
\end{align}

In order to obtain $\delta k^{\mu}$, one can use  the normalization and
orthogonality relations to deduce
\begin{align}
\delta\left(k^{\mu}k_{\mu}\right) & \;=\;0\;,\\
\delta\left(u^{\mu}k_{\mu}\right) & \;=\;0\;,
\end{align}
from which it follows that
\begin{align}
k^{\mu}\delta k_{\mu} & \;=\;-\delta k^{\mu}k_{\mu}\;,
\end{align}
or, more explicitly,
\begin{equation}
k^{r}\delta k_{r}\;=\;-\delta k^{r}k_{r}-\delta k^{t}k_{t}\;.\label{eq:kk}
\end{equation}

One can now use the knowledge that
\begin{align}
\delta k_{r} & \;=\;\delta\left(k^{\mu}g_{\mu r}\right)\nonumber \\
 & \;=\;\delta k^{\mu}g_{\mu r}+k^{\mu}\delta g_{\mu r}\nonumber \\
 & \;=\;\delta k^{r}g_{rr}+\delta k^{t}g_{tr}+k^{r}h_{rr}\;,\label{eq:deltak_r}
\end{align}
in order to rewrite Eq.~(\ref{eq:kk}) as
\begin{equation}
\delta k^{r}\left(k^{r}g_{rr}+k_{r}\right)+\left(k^{r}\right)^{2}h_{rr}\;=\;-\delta k^{t}\left(k^{r}g_{tr}+k_{t}\right)\;.
\end{equation}
The previous equation can be further rewritten by using $\;k_{r}=k^{r}g_{rr}\;$ and $\;k_{t}=g_{tr}k^{r}\;$ to arrive at
\begin{equation}
2\delta k^{r}g_{rr}+k^{r}h_{rr}\;=\;-2\delta k^{t}g_{tr}\;.
\label{eq:kr}
\end{equation}

From the orthogonality of $u^{\mu}$ and $k_{\mu}$, it follows that
\begin{align}
0 & \;=\;\delta\left(u^{\mu}k_{\mu}\right)\nonumber \\
 & \;=\;\delta u^{\mu}k_{\mu}+u^{\mu}\delta k_{\mu}\nonumber \\
 & \;=\;\delta u^{t}k_{t}+\delta u^{r}k_{r}+u^{t}\delta k_{t}\;,
\end{align}
which then yields
\begin{align}
\delta k_{t} & \;=\;-\frac{\delta u^{r}}{u^{t}}k_{r}\nonumber \\
 & \;=\;-k_{r}\partial_{t}\xi^{r}\nonumber \\
 & \;=\;-\partial_{t}\left(\frac{e^{-\Lambda}W}{r^{2}}e^{i\omega t}Y_{\ell m}\right)e^{\Lambda}\left[1-\frac{1}{2}H_{0}e^{i\omega t}Y_{\ell m}\right]\nonumber \\
 & \;\approx\;-i\omega\frac{W}{r^{2}}e^{i\omega t}Y_{\ell m}\;.\label{eq:deltak_t}
\end{align}

One can also use
\begin{align}
\delta k^{t} & \;=\;\delta\left(g^{t\mu}k_{\mu}\right)\nonumber \\
 & \;=\;\delta g^{t\mu}k_{\mu}+g^{t\mu}\delta k_{\mu}\nonumber \\
 & \;=\;-h^{tt}k_{t}+g^{tt}\delta k_{t}-h^{tr}k_{r}+h^{tr}\delta k_{r}\;
\end{align}
and then substitute for  $\delta k_{t},\delta k_{r}$, with their respective
expressions from Eqs. (\ref{eq:deltak_t}) and (\ref{eq:deltak_r}), to
obtain
\begin{equation}
\delta k^{t}\left(1-h^{tr}h_{tr}\right)\;=\;-h^{tt}k_{t}+g^{tt}\delta k_{t}-h^{tr}k_{r}+h^{tr}\delta k^{r}g_{rr}+h^{tr}k^{r}h_{rr}\;. \label{katy}
\end{equation}

In this way,  one now possesses a pair of equations, (\ref{eq:kr}) and (\ref{katy}), for $\delta k^{t}$
and $\delta k^{r}$, with solutions
\begin{equation}
\delta k^{r}\;=\;\frac{1}{2}e^{-\Lambda}H_{0}e^{i\omega t}Y_{\ell m}\ ,
\end{equation}
and
\begin{equation}
\delta k^{t}\;\approx\; i\omega e^{-2\Phi}\left[r^{-2}W-e^{-\Lambda}H_{1}\right]e^{i\omega t}Y_{\ell m}\;.
\end{equation}

From Eq.~(\ref{eq:deltak_r}) and the above solutions, it now follows that
\begin{equation}
\delta k_{r}\;=\;-\frac{1}{2}e^{\Lambda}H_{0}e^{i\omega t}Y_{\ell m}\;.
\end{equation}

To derive $\delta k_{\theta,\phi}$, one requires the use of the Einstein equations. In particular,
\begin{align}
\delta G_{\phi}^{\phantom{r}r} &\; =\; \frac{1}{2}e^{-2\left(\Lambda+\Phi\right)}e^{i\omega t}\partial_{\phi}Y_{\ell m}\left[-\omega^{2}H_{1}+ e^{2\Phi}\left(-2\Phi^{\prime}H_{0}-H_{0}^{\prime}+ K^{\prime}\right)\right]\nonumber \\
 & \;=\; \delta T_{\phi}^{\;\;r}\;=\;\left(p-q\right)\delta k_{\phi}k^{r}\ ,
\end{align}
and
\begin{align}
\delta G_{\theta}^{\phantom{r}r} & \;=\;\frac{1}{2}e^{-2\left(\Lambda+\Phi\right)}e^{i\omega t}\partial_{\theta}Y_{\ell m} \left[-\omega^{2}H_{1}+e^{2\Phi}\left(-2\Phi^{\prime}H_{0}-H_{0}^{\prime} +K^{\prime}\right)\right]\nonumber \\
 & \;=\;  \delta T_{\theta}^{\;\;r}\;=\;\left(p-q\right)\delta k_{\theta}k^{r}\;,
\end{align}
from which it follows that
\begin{equation}
\widehat{\delta k_{\phi}}\;=\;\widehat{\delta k_{\theta}}\;,
\end{equation}
where a hat denotes a quantity stripped of its angular dependence ({\em i.e.},
stripped of its spherical harmonic or derivative thereof), so that
$\;
\delta k_{\phi}\left(r,\theta,\phi\right)=\widehat{\delta k_{\phi}}\left(r\right) \partial_{\phi}Y_{\ell m}\left(\theta,\phi\right)\;$ and
$\;\delta k_{\theta}\left(r,\theta,\phi\right)=\widehat{\delta k_{\theta}}\left(r\right) \partial_{\theta}Y_{\ell m}\left(\theta,\phi\right)\;$.
Let us remind the reader that, unlike the choice
made in \cite{mondal2023nonradial}, $\delta k_{\theta,\phi}$ are allowed to be non-vanishing, which directly affects the resulting Einstein equations (see Section~A.4).

\subsection{Matter perturbations}

In the following, the perturbed energy-conservation equation,
\begin{equation}
\delta\left(\nabla_{\nu}T_{\phantom{\mu}\mu}^{\nu}\right)\;=\;0\;,
\end{equation}
is used to obtain the variations of matter-related quantities, such
as the energy density $\delta\rho$ and the  pressure components $\delta p,\delta q$.

Beginning
with the EM tensor as defined in Eq.~(\ref{eq:SET}), one obtains
the zeroth-order conservation equation,
\begin{equation}
\nabla_{\nu}T_{\phantom{\mu}\mu}^{\nu}\; =\;\nabla_{\nu}\left(\left(\rho+q\right)u_{\mu}u^{\nu}\right) +\nabla_{\nu}\left(\left(p-q\right)k_{\mu}k^{\nu}\right)+\nabla_{\mu}q\;=\;0\;.
\end{equation}
Its variation, when projected along the direction of the velocity
$u^{\mu}$, is then given by (with $\;\sigma=p-q$)
\begin{align}
0 & \;=\;u^{\mu}\delta\left(\nabla_{\nu}T_{\mu}^{\nu}\right)\nonumber \\
 & \;=\;-u^{\nu}\nabla_{\nu}\delta\rho-\nabla_{\nu}\left\{ \left[\left(\rho+q\right)\delta_{\mu}^{\nu}+\sigma k_{\mu}k^{\nu}\right]\delta u^{\mu}\right\} -\left(\rho+q\right)a_{\mu}\delta u^{\mu}-\left(\nabla_{\nu}u^{\mu}\right)\delta\left(\sigma k_{\mu}k^{\nu}\right)\nonumber \\
 & -\left(\rho+q\right)\delta\Gamma_{\alpha\nu}^{\nu}u^{\alpha}-u^{\alpha}\sigma k_{\mu}k^{\nu}\delta\Gamma_{\alpha\nu}^{\mu} +\frac{1}{2}h_{\mu\alpha}u^{\mu}u^{\alpha}u^{\nu}\nabla_{\nu}\left(\rho+q\right)\;.
 \label{eq:udelTprep}
\end{align}

In deriving Eq.~(\ref{eq:udelTprep}), we have used that, for any vector $V_{\mu}$,
\begin{align}
\delta\left(\nabla_{\nu}V_{\mu}\right) & \;=\;\delta\left(\partial_{\nu}V_{\mu}-\Gamma_{\mu\nu}^{\alpha}V_{\alpha}\right)\nonumber \\
 & \;=\;\partial_{\nu}\delta V_{\mu}-\delta\left(\Gamma_{\mu\nu}^{\alpha}V_{\alpha}\right)\nonumber \\
 & \;=\;\partial_{\nu}\delta V_{\mu}-\Gamma_{\mu\nu}^{\alpha}\delta V_{\alpha}-\delta\Gamma_{\mu\nu}^{\alpha}V_{\alpha}\nonumber \\
 & \;=\;\nabla_{\nu}\delta V_{\mu}-\delta\Gamma_{\mu\nu}^{\alpha}V_{\alpha}\;,
\end{align}
and that the velocity satisfies $\;u^{\mu}\nabla_{\nu}u_{\mu}=0\;$, as
well as both
\begin{align}
0\;=\;\delta\left(\nabla_{\nu}u^{\nu}\right) & \;=\;\nabla_{\nu}\delta u^{\nu}+\delta\Gamma_{\alpha\nu}^{\nu}u^{\alpha}\nonumber \\
\;\implies\;\nabla_{\nu}\delta u^{\nu} & \;=\;-\delta\Gamma_{\alpha\nu}^{\nu}u^{\alpha}\;,
\end{align}
and
\begin{align*}
\delta\left(u^{\mu}u_{\mu}\right) & \;=\;0\\
 & \;=\;\delta u^{\mu}u_{\mu}+u^{\mu}\delta u_{\mu}\\
 & \;=\;\delta u^{\mu}u_{\mu}+u^{\mu}\delta\left(g_{\mu\alpha}u^{\alpha}\right)\\
 & \;=\;\delta u^{\mu}u_{\mu}+u^{\mu}h_{\mu\alpha}u^{\alpha}+u_{\alpha}\delta u^{\alpha}\\
\;\implies\;\delta u^{\mu}u_{\mu} & \;=\;-\frac{1}{2}h_{\mu\alpha}u^{\mu}u^{\alpha}\;.
\end{align*}

One  is then able to use Eq.~(\ref{eq:udelTprep}) so as to  isolate the derivative of $\delta\rho$
(and recall
that $\;u^{\mu}=u^{t}\delta_{t}^{\mu}\;$ while $\;k^{\mu}=k^{r}\delta_{r}^{\mu}\;$),
\begin{align}
u^{t}\partial_{t}\delta\rho & \;=\;-\nabla_{\nu}\left\{ \left[\left(\rho+q\right)\delta_{\mu}^{\nu}+\sigma k_{\mu}k^{\nu}\right]\delta u^{\mu}\right\} -\left(\rho+q\right)a_{\mu}\delta u^{\mu}-\left(\nabla_{\nu}u^{\mu}\right)\delta\left(\sigma k_{\mu}k^{\nu}\right)\nonumber \\
& -\left(\rho+q\right)\delta\Gamma_{t\nu}^{\nu}u^{t}-u^{t}\sigma k_{r}k^{r}\delta\Gamma_{tr}^{r}+ \frac{1}{2}h_{tt}\left(u^{t}\right)^{3}\underset{\;=\;0}{\underbrace{\partial_{t}\left(\rho+q\right)}} \nonumber \\
& \;=\;-\nabla_{j}\left\{ \left[\left(\rho+q\right)\delta_{i}^{j}+\sigma k_{i}k^{j}\right]u^{t}\partial_{t}\xi^{i}\right\} -u^{t}\left(\rho+q\right)a_{r}\partial_{t}\xi^{r}- \left(\partial_{\nu}u^{\mu}+\Gamma_{\alpha\nu}^{\mu}u^{\alpha}\right)\delta\left(\sigma k_{\mu}k^{\nu}\right)\nonumber \\
 & -\left(\rho+q\right)\left[\delta\Gamma_{tt}^{t}+ \delta\Gamma_{tr}^{r}+\delta\Gamma_{t\theta}^{\theta} +\delta\Gamma_{t\phi}^{\phi}\right]u^{t}-\sigma u^{t}\delta\Gamma_{tr}^{r}
 \nonumber \\
 & \;=\;-\nabla_{j}\left\{ \left[\left(\rho+q\right)\delta_{i}^{j}+\sigma k_{i}k^{j}\right]u^{t}\partial_{t}\xi^{i}\right\} -u^{t}\left(\rho+q\right)a_{r}\partial_{t}\xi^{r}\nonumber \\
 & -\underset{\sim\mathcal{O}\left(\delta^{2}\right)}{\underbrace{\delta\sigma k_{t}}} \partial_{r}u^{t}k^{r}-\underset{\sim\mathcal{O}\left(\delta^{2}\right)} {\underbrace{\delta\sigma k_{t}}}\Gamma_{tr}^{t}u^{t}k^{r} -\underset{\sim\mathcal{O}\left(\delta^{2}\right)} {\underbrace{\delta\sigma\Gamma_{tr}^{r}}}u^{t}k_{r}k^{r}
 -\sigma k^{r}\delta k_{t}\partial_{r}u^{t}\nonumber \\
 & -\sigma k^{r}u^{t}\left[\Gamma_{tr}^{t}\delta k_{t}+\Gamma_{tr}^{r}\delta k_{r}+\Gamma_{tr}^{\theta}\delta k_{\theta}+\Gamma_{tr}^{\phi}\delta k_{\phi}\right]-\sigma\partial_{\nu}u^{t} \underset{\sim\mathcal{O}\left(\delta^{2}\right)} {\underbrace{k_{t}\delta k^{\nu}}}-\sigma u^{t}\left[\Gamma_{t\nu}^{t} \underset{\sim\mathcal{O}\left(\delta^{2}\right)}{\underbrace{k_{t}\delta k^{\nu}}}+\Gamma_{t\nu}^{r}k_{r}\delta k^{\nu}\right]\nonumber \\
 & -\left(\rho+q\right)\left[\delta\Gamma_{tt}^{t}+ \delta\Gamma_{tr}^{r}+\delta\Gamma_{t\theta}^{\theta} \delta\Gamma_{t\phi}^{\phi}\right]u^{t}-\sigma u^{t}k^{r}\left(\underset{\sim\mathcal{O}\left(\delta^{2}\right)} {\underbrace{k_{t}\delta\Gamma_{tr}^{t}}}+k_{r}\delta\Gamma_{tr}^{r}\right)\;,
\end{align}
where it was also used that $\rho$ and $q$ are stationary background
quantities. Discarding all higher-order expressions and dividing by
$u^{t}$, one then has
\begin{align}
\partial_{t}\delta\rho & \;=\;-\frac{1}{u^{t}}\nabla_{j}\left\{ \left[\left(\rho+q\right)\delta_{i}^{j}+\sigma k_{i}k^{j}\right]u^{t}\partial_{t}\xi^{i}\right\} -\left(\rho+q\right)a_{r}\partial_{t}\xi^{r}\nonumber \\
 & -\sigma\left[k^{r}\Gamma_{tr}^{t}\delta k_{t}+\underset{\sim\mathcal{O}\left(\delta^{2}\right)}{\underbrace{k^{r}\Gamma_{tr}^{r}\delta k_{r}+k^{r}\Gamma_{tr}^{\theta}\delta k_{\theta}+k^{r}\Gamma_{tr}^{\phi}\delta k_{\phi}}}+\delta k_{t}k^{r}\partial_{r}\left(\ln u^{t}\right)+\Gamma_{t\nu}^{r}k_{r}\delta k^{\nu}+\delta\Gamma_{tr}^{r}\right]\nonumber \\
 & -\left(\rho+q\right)\left[\delta\Gamma_{tt}^{t}+\delta\Gamma_{tr}^{r} +\delta\Gamma_{t\theta}^{\theta}+\delta\Gamma_{t\phi}^{\phi}\right]\;.
\end{align}

Let us next rewrite the previous equation whereby  all spacetime perturbations
are expressed explicitly  in terms of the  Regge--Wheeler gauge ({\em cf},
Eqs. (\ref{evenmetric}) and (\ref{eq:xievendef})) such that
\begin{align}
\partial_{t}\delta\rho & \;=\;-\frac{1}{u^{t}}\nabla_{j}\left\{ \left[\left(\rho+q\right)\delta_{i}^{j}+\sigma k_{i}k^{j}\right]u^{t}\partial_{t}\xi^{i}\right\} -\left(\rho+q\right)a_{r}\partial_{t}\xi^{r}\nonumber \\
 & -\sigma\left[k_{r}\left[\Gamma_{tt}^{r}\delta k^{t}+\underset{\sim\mathcal{O}\left(\delta^{2}\right)}{\underbrace{\Gamma_{tr}^{r}\delta k^{r}}}+\underset{\sim\mathcal{O}\left(\delta^{2}\right)}{\underbrace{\Gamma_{t\theta}^{r}\delta k^{\theta}}}\right]+\delta\Gamma_{tr}^{r}+k^{r}\delta k_{t}\left[\partial_{r}\left(\ln u^{t}\right)+\Gamma_{tr}^{t}\right]\right]\nonumber \\
 & -\left(\rho+q\right)\left[\delta\Gamma_{tt}^{t}+\delta\Gamma_{tr}^{r}+ \delta\Gamma_{t\theta}^{\theta}+\delta\Gamma_{t\phi}^{\phi}\right]\nonumber \\
 & \;=\;-\frac{i\omega}{u^{t}}\nabla_{j}\left\{ \left[\left(\rho+q\right)\delta_{i}^{j}+\sigma k_{i}k^{j}\right]u^{t}\xi^{i}\right\} -\left(\rho+q\right)a_{r}\partial_{t}\xi^{r}\nonumber \\
 & -\underset{\;=\;\sigma a_{r}\partial_{t}\xi^{r}}{\underbrace{i\omega\sigma\Phi^{\prime}r^{-2}e^{-\Lambda}We^{i\omega t}Y_{\ell m}}}+\sigma\frac{1}{2}i\omega H_{0}e^{i\omega t}Y_{\ell m}+i\omega\left(\rho+q\right)Ke^{i\omega t}Y_{\ell m}\;,
 \label{derivrho}
\end{align}
where we have used that
\begin{align}
a_{r} & \;=\;u^{t}\left(\partial_{t}u_{r}-\Gamma_{rt}^{t}u_{t}-\Gamma_{rt}^{r}u_{r}\right)\nonumber \\
 & \;=\;\Phi^{\prime}+\mathcal{O}\left(\delta\right)\;,
\end{align}
as well as both of
\begin{align}
a_{\theta} & \;=\;u^{t}\left(\partial_{t}u_{\theta}-\Gamma_{\theta t}^{t}u_{t}-\Gamma_{\theta t}^{r}u_{r}\right)\nonumber \\
 & \;=\;u^{t}\left(\frac{1}{2}H_{0}u_{t}+\frac{1}{2}i\omega e^{-2\Lambda}H_{1}u_{r}\right)\partial_{\theta}Y_{\ell m}\nonumber \\
 & \;=\;-\frac{1}{2}H_{0}e^{i\omega t}\partial_{\theta}Y_{\ell m}+\mathcal{O}\left(\delta^{2}\right),
\end{align}
and
\begin{align}
a_{\phi} & \;=\;u^{t}\left(\partial_{t}u_{\phi}-\Gamma_{\phi t}^{t}u_{t}-\Gamma_{\phi t}^{r}u_{r}\right)\nonumber \\
 & \;=\;u^{t}\left(\frac{1}{2}H_{0}u_{t}+\frac{1}{2}i\omega e^{-2\Lambda}H_{1}u_{r}\right)\partial_{\phi}Y_{\ell m}\nonumber \\
 & \;=\;-\frac{1}{2}H_{0}e^{i\omega t}\partial_{\phi}Y_{\ell m}+\mathcal{O}\left(\delta^{2}\right)\;.
\end{align}

Integrating  Eq.~(\ref{derivrho}) and thus $\partial_{t}\rho$ over time, one obtains that, to leading order,
\begin{align}
\delta\rho & \;=\;-\frac{1}{\sqrt{-g}}\partial_{j}\left\{ \sqrt{-g}\left[\left(\rho+q\right)\delta_{i}^{j}+\sigma k_{i}k^{j}\right]\xi^{i}\right\} \nonumber \\
 & -\left\{ \left[\left(\rho+q\right)\delta_{i}^{j}+\sigma k_{i}k^{j}\right]\xi^{i}\right\} \nabla_{j}\left(\ln u^{t}\right)-\left(\rho+p\right)a_{r}\xi^{r}\nonumber \\
 & +\frac{1}{2}\sigma H_{0}e^{i\omega t}Y_{\ell m}+\left(\rho+q\right)Ke^{i\omega t}Y_{\ell m}\ ,
\end{align}
or, more explicitly (but again to leading order),
\begin{align}
\delta\rho & \;=\;\left[-\left(\Lambda^{\prime}+\Phi^{\prime}+\frac{2}{r}\right) \left(\rho+p\right)-\partial_{r}\left(\rho+p\right)\right]\frac{e^{-\Lambda}W}{r^{2}}e^{i\omega t}Y_{\ell m}-\left(\rho+p\right)\partial_{r}\left(\frac{e^{-\Lambda}W}{r^{2}}\right)e^{i\omega t}Y_{\ell m}\nonumber \\
 & +\left(\rho+q\right)\frac{V}{r^{2}}e^{i\omega t}\underset{\;=\;-\ell\left(\ell+1\right)Y_{\ell m}}{\underbrace{\left\{ \frac{1}{\sin\theta}\partial_{\theta}\left(\sin\theta\partial_{\theta}Y_{\ell m}\right)+\frac{1}{\sin^{2}\theta}\partial_{\phi}^{2}Y_{\ell m}\right\} }}\nonumber \\
 & -\left\{ -\Phi^{\prime}\left(\rho+p\right)\frac{e^{-\Lambda}W}{r^{2}}e^{i\omega t}Y_{\ell m}+\left(\rho+q\right)\underset{\sim\mathcal{O}\left(\delta^{2}\right)} {\underbrace{\xi^{\theta}\nabla_{\theta}\left(\ln u^{t}\right)}}\right\} -\left(\rho+p\right)\Phi^{\prime}\frac{e^{-\Lambda}W}{r^{2}}e^{i\omega t}Y_{\ell m}
 \nonumber \\
 & +\frac{1}{2}\sigma H_{0}e^{i\omega t}Y_{\ell m}+\left(\rho+q\right)Ke^{i\omega t}Y_{\ell m}\nonumber \\
 & \;=\;\left\{ -\left(\rho+p\right)\left[\frac{e^{-\Lambda}W^{\prime}}{r^{2}} +\frac{\ell\left(\ell+1\right)}{r^{2}}V\right] -\frac{d\rho}{dr}e^{-\Lambda}\frac{W}{r^{2}}+\frac{2\sigma}{r^{3}}e^{-\Lambda}W +\sigma\frac{\ell\left(\ell+1\right)}{r^{2}}V\right\} e^{i\omega t}Y_{\ell m}
 \nonumber \\
 & +\frac{1}{2}\sigma H_{0}e^{i\omega t}Y_{\ell m}+\left(\rho+q\right)Ke^{i\omega t}Y_{\ell m}\;,\label{eq:deltarho}
\end{align}
where the zeroth-order energy-conservation equation,
\begin{equation}
\frac{dp}{dr}\;=\;-\Phi^{\prime}\left(p+\rho\right)-\frac{2}{r}\sigma\;,
\end{equation}
has been applied.

One may now obtain $\delta p$ from $\delta \rho$ by using the relation
between the Eulerian and Lagrangian variations of the radial pressure ($\delta p$ and $\Delta p$,
respectively),
\begin{align}
\delta p & \;=\;\Delta p-\xi^{r}\partial_{r}p\nonumber \\
 & \;=\;\frac{dp}{d\rho}\Delta\rho-\xi^{r}\partial_{r}p\nonumber \\
 & \;=\;\frac{dp}{d\rho}\left(\delta\rho+\xi^{r}\partial_{r}\rho\right)-\xi^{r}\partial_{r}p
 \nonumber \\
 & \;=\;\frac{dp}{d\rho}\delta\rho\;.
\end{align}
Therefore, the variation of the radial pressure is, to leading order,
\begin{align}
\delta p & \;=\;\frac{dp}{d\rho}\left\{ -\left(\rho+p\right)\left[\frac{e^{-\Lambda}W^{\prime}}{r^{2}} +\frac{\ell\left(\ell+1\right)}{r^{2}}V\right] +\frac{2\sigma}{r^{3}}e^{-\Lambda}W+\sigma\frac{\ell\left(\ell+1\right)}{r^{2}}V\right\} e^{i\omega t}Y_{\ell m}\nonumber \\
 & -\frac{dp}{dr}e^{-\Lambda}\frac{W}{r^{2}}e^{i\omega t}Y_{\ell m}+\frac{dp}{d\rho}\left[\frac{1}{2}\sigma H_{0}e^{i\omega t}Y_{\ell m}+\left(\rho+q\right)Ke^{i\omega t}Y_{\ell m}\right]\;.
 \label{eq:deltap}
\end{align}

By the same token, $\;\delta q=\left(\partial q/\partial p\right)\delta p\;$, so that
\begin{align}
\delta q & \;=\;\frac{dq}{d\rho}\left\{ -\left(\rho+p\right)\left[\frac{e^{-\Lambda}W^{\prime}}{r^{2}} +\frac{\ell\left(\ell+1\right)}{r^{2}}V\right] +\frac{2\sigma}{r^{3}}e^{-\Lambda}W+\sigma\frac{\ell\left(\ell+1\right)}{r^{2}}V\right\} e^{i\omega t}Y_{\ell m}
\nonumber \\
 & -\frac{dq}{dr}e^{-\Lambda}\frac{W}{r^{2}}e^{i\omega t}Y_{\ell m}+\frac{dq}{d\rho} \left[\frac{1}{2}\sigma H_{0}e^{i\omega t}Y_{\ell m}+
 \left(\rho+q\right)Ke^{i\omega t}Y_{\ell m}\right]\;.
\end{align}

\subsection{The perturbed Einstein equations\label{subsec:Perturbed-Einstein-equations}}

Let us recall our convention that $\; 8 \pi G=1\;$.

We begin   here  with  the identity $\;\delta G_{\phantom{\theta}\theta}^{\theta}-\delta G_{\phantom{\phi}\phi}^{\phi}= \delta T_{\phantom{\theta}\theta}^{\theta}-\delta T_{\phantom{\phi}\phi}^{\phi} =0\;$,
which is, quite simply,
\begin{equation}
H_{0}-H_{2}\;=\;0\;.
\end{equation}
This relationship will be  used to eliminate
$H_{2}$ from all subsequent equations.

Let us next  consider
$\;\delta G_{\phantom{t}t}^{t}=\delta T_{\phantom{t}t}^{t}=-\delta\rho\;$,
which yields
\begin{align}
-\delta\rho
& \;=\;-\frac{1}{2r^{2}}e^{i\omega t}Y_{\ell m}\left\{ 2K-\ell\left(\ell+1\right)\left(H_{0}+K\right)\right.\nonumber \\
 & \left.+2e^{-2\Lambda}\left[H_{0}\left(2r\Lambda^{\prime}-1\right)-r\left(H_{0}^{\prime} +\left(r\Lambda^{\prime}-3\right)K^{\prime}-rK^{\prime\prime}\right)\right]\right\}\;.\label{eq:H0'withdeltarho}
\end{align}
When the above  is  combined with Eq.~(\ref{eq:deltarho}), one finds that
\begin{align}
0 & \;=\;H_{0}^{\prime}+r^{-1}e^{2\Lambda}\left[1-r^{2}\rho +\frac{\ell\left(\ell+1\right)}{2}+\frac{\sigma}{2}r^{2}\right]H_{0}\nonumber \\
 & -e^{2\Lambda}\left(3-\frac{5m\left( r\right)}{8 \pi r} -\frac{r^{2}\rho}{2}\right)K^{\prime}-rK^{\prime\prime} +r^{-1}e^{2\Lambda}\left[\frac{\ell\left(\ell+1\right)}{2}-1 +r^{2}\left(\rho+q\right)\right]K\nonumber \\
 & +r^{-1}\left[\frac{2\sigma}{r}-\frac{d\rho}{dr}\right]e^{\Lambda}W -r^{-1}e^{2\Lambda}\left(\rho+q\right)\ell\left(\ell+1\right)V -r^{-1}\left(\rho+p\right)e^{\Lambda}W^{\prime}\;.
 \label{eq:H0'explicit}
\end{align}

The third equation of interest is $\;\delta G_{\phantom{t}t}^{r}= \delta T_{\phantom{0}t}^{r}\;$, for which  the relevant tensors,
\begin{align}
\delta G^{r}_{\phantom{t}t} & \;=\;-i\omega\frac{e^{-2\Lambda}}{2r^{2}}\left\{ \ell\left(\ell+1\right)H_{1}+2r\left[H_{0} +\left(r\Phi^{\prime}-1\right)K-rK^{\prime}\right]\right\} e^{i\omega t}Y_{\ell m}\ ,
\end{align}
and
\begin{align}
\delta T^{r}_{\phantom{t}t} & \;=\;\left(\rho+q\right)u_{t}\delta u^{r}+\left(p-q\right)\delta k_{t}k^{r}\nonumber \\
 & \;=\;-\left(\rho+p\right)r^{-2}e^{-\Lambda}\partial_{t}WY_{\ell m}\nonumber \\
 & \;=\;-i\omega\left(\rho+p\right)r^{-2}e^{-\Lambda}We^{i\omega t}Y_{\ell m}\;,
\end{align}
   can be combined into
\begin{equation}
\ell\left(\ell+1\right)H_{1}\;=\;-2r\left[H_{0}+\left(r\Phi^{\prime}-1\right)K -rK^{\prime}\right]+2\left(\rho+p\right)e^{\Lambda}W\;.
\end{equation}

The last such initial-value equation to consider  is $\;\delta G^{r}_{\phantom{\theta}\theta}=\delta T^{r}_{\phantom{\theta}\theta}\;$, where
\begin{align}
\delta G^{r}_{\phantom{\theta}\theta} & \;=\;-\frac{e^{-2\left(\Lambda+\Phi\right)}}{2}e^{i\omega t}\partial_{\theta}Y_{\ell m}\left[\omega^{2}H_{1}+e^{2\Phi}\left(2\Phi^{\prime}H_{0} +H_{0}^{\prime}-K^{\prime}\right)\right]\;,
\label{eq:deltaGthetar}
\end{align}
and
\begin{align}
\delta T^{r}_{\phantom{\theta}\theta} & \;=\;\left(p-q\right)\delta k_{\theta}k^{r}\;.
\end{align}

Continuing with the other non-vanishing members of  the Einstein equations,  we can  use $\;\delta G^t_{\;\;\phi}=\delta T^t_{\;\;\phi}\;$
to  relate
\begin{align}
\delta T^t_{\;\;\phi}\;=\;\left(\rho+q\right)\delta u_{\phi}u^{t}\;,
\end{align}
with
\begin{align}
G^t_{\;\;\phi}& \;=\;-\frac{i\omega}{2}e^{-2\left(\Lambda+\Phi\right)}e^{i\omega t}\partial_{\phi}Y_{\ell m}\left[H_{1}\left(\Lambda^{\prime}-\Phi^{\prime}\right) +e^{2\Lambda}\left(H_{0}+K\right)-H_{1}^{\prime}\right]\nonumber \\
 & \;=\;\left(\rho+q\right)\delta u_{\phi}u^{t}\nonumber \\
 & \;=\;-i\omega\left(\rho+q\right)e^{-2\Phi}Ve^{i\omega t}\partial_{\phi}Y_{\ell m}\;,
\end{align}
as well as apply
\begin{align}
\Lambda^{\prime}-\Phi^{\prime} & \;=\;r^{-1}e^{2\Lambda}\left(\frac{1}{2}r^{2}\left(\rho-p\right)+e^{-2\Lambda}-1\right)
\nonumber \\
 & \;=\;-r^{-1}e^{2\Lambda}\left(\frac{1}{2}r^{2}\left(p-\rho\right) +\frac{m(r)} {4\pi r}\right)\;,
\end{align}
so as to obtain
\begin{equation}
H_{1}^{\prime}\;=\;-r^{-1}e^{2\Lambda}\left(\frac{1}{2}r^{2}\left(p-\rho\right) +\frac{m\left( r\right)}{4\pi r}\right)H_{1} +e^{2\Lambda}\left(H_{0}+K-2\left(\rho+q\right)V\right)\;.
\label{eq:H1'appendix}
\end{equation}

The equation $\;\delta G^r_{\;\;r}=\delta T^r_{\;\;r}\;$
plus the result
\begin{align}
\delta T^{r}_{\;\;r} & \;=\;\delta p+\left(\rho+q\right)u_{r}\delta u^{r}-\left(p-q\right)\left(\delta k^{t}k_{t}\right)\nonumber \\
 & \;=\;\delta p+\mathcal{O}\left(\delta^{2}\right)\ ,
\end{align}
amounts to
\begin{align}
\delta p\;=\;\delta G_{r}^{\phantom{r}r}\;=\; & \frac{e^{-2\left(\Lambda+\Phi\right)}}{8 r^{2}}e^{i\omega t}Y_{\ell m}\left\{ 4r\omega^{2}H_{1}-e^{2\Lambda}\left(2r^{2}\omega^{2} +e^{2\Phi}\left(2-\ell\left(\ell+1\right)\right)\right)K\right.\nonumber \\
 & \left.+2e^{2\Phi}\left[rH_{0}^{\prime}-r\left(1+r\Phi^{\prime}\right)K^{\prime}\right] +\left[-e^{2\left(\Lambda+\Phi\right)}\ell\left(\ell+1\right) +2e^{2\Phi}\left(1+2r\Phi^{\prime}\right)\right]H_{0}\right\}\;.
 \label{eq:K''withdeltap}
\end{align}
Substituting for $\delta p$ via Eq.~(\ref{eq:deltap})
and  for $H_{0}^{\prime}$ via Eq.~(\ref{eq:H0'explicit}), we then end up with
\begin{align}
e^{-2\Phi}\omega^{2}K & -e^{-2\Lambda}K^{\prime\prime}-2r^{-1}\left(e^{-2\Lambda} -\frac{r^{2}\left(\rho+p\right)}{4}\right)K^{\prime}\nonumber \\
 & -\left[r^{2}\left(\rho+p\right)-\ell\left(\ell+1\right) -\frac{\sigma}{2}r^{2} \left(1+\frac{dp}{d\rho}\right)\right] \frac{H_{0}}{r^{2}}+r^{2}\left(\rho+q\right) \left(1+\frac{dp}{d\rho}\right)\frac{K}{r^{2}}\nonumber \\
 & +\left(1+\frac{dp}{d\rho}\right)\left[\frac{2\sigma}{r} -\frac{d\rho}{dr}\right]\frac{e^{-\Lambda}W}{r^{2}} -\left(1+\frac{dp}{d\rho}\right)\left(\rho+p\right)\frac{e^{-\Lambda}W^{\prime}}{r^{2}}
 \nonumber \\
 & -\left(1+\frac{dp}{d\rho}\right)\left(\rho+q\right)\ell\left(\ell+1\right)\frac{V}{r^{2}} -2\omega^{2}e^{-2\left(\Lambda+\Phi\right)}\frac{H_{1}}{r}\nonumber \\
 & \;=\;0\;.
\end{align}

Let us now pivot to the variation of the energy-conservation equation,
\begin{align}
0\;=\;\delta\left(T_{\mu\phantom{;\nu};\nu}^{\phantom{\nu}\nu}\right) & \;=\;\partial_{\nu}\left(\delta\rho+\delta q\right)u_{\mu}u^{\nu}+\left(\delta\rho+\delta q\right)a_{\mu}\nonumber \\
 & +\partial_{\nu}\left(\rho+q\right)\delta\left(u_{\mu}u^{\nu}\right) +\left(\rho+q\right)\delta\left(\nabla_{\nu}u_{\mu}u^{\nu}\right)\nonumber \\
 & +\partial_{\nu}\delta\sigma k_{\mu}k^{\nu}+\partial_{\nu}\sigma\delta k_{\mu}k^{\nu}+\partial_{\nu}\sigma k_{\mu}\delta k^{\nu}\nonumber \\
 & +\delta\sigma\nabla_{\nu}k_{\mu}k^{\nu} +\sigma\delta\left(\nabla_{\nu}k_{\mu}\right)k^{\nu}+\sigma\nabla_{\nu}k_{\mu}\delta k^{\nu}\nonumber \\
 & +\delta\sigma k_{\mu}\nabla_{\nu}k^{\nu} +\sigma\delta k_{\mu}\nabla_{\nu}k^{\nu}+\sigma k_{\mu}\delta\left(\nabla_{\nu}k^{\nu}\right)\nonumber \\
 & +\partial_{\mu}\delta q\;.
\end{align}

Choosing the free index to be $r$, we then have
\begin{align}
0\;=\;\delta\left(T_{r\phantom{;\nu};\nu}^{\phantom{\nu}\nu}\right) & \;=\;\partial_{\nu}\left(\delta\rho+\delta q\right)u_{r}u^{\nu}+\left(\delta\rho+\delta q\right)a_{r}+\partial_{\nu}\left(\rho+q\right)\delta\left(u_{r}u^{\nu}\right)\nonumber \\
 & +\left(\rho+q\right)\delta\left(\nabla_{\nu}u_{r}u^{\nu}\right)+\partial_{\nu}\delta\sigma k_{r}k^{\nu}+\partial_{\nu}\sigma\delta k_{r}k^{\nu}+\partial_{\nu}\sigma k_{r}\delta k^{\nu}\nonumber \\
 & +\delta\sigma\nabla_{\nu}k_{r}k^{\nu} +\sigma\delta\left(\nabla_{\nu}k_{r}\right)k^{\nu}+\sigma\nabla_{\nu}k_{r}\delta k^{\nu}\nonumber \\
 & +\delta\sigma k_{r}\nabla_{\nu}k^{\nu}+\sigma\delta k_{r}\nabla_{\nu}k^{\nu}+\sigma k_{r}\delta\left(\nabla_{\nu}k^{\nu}\right)+\partial_{r}\delta q\nonumber \\
 & \;=\;u^{t}\overset{\sim\mathcal{O}\left(\delta^{2}\right)} {\overbrace{u_{r}\partial_{t}\left(\delta\rho+\delta q\right)}}+\left(\delta\rho+\delta q\right)a_{r}\nonumber \\
 & +\delta u_{r}u^{t}\overset{\;=\;0}{\overbrace{\partial_{t}\left(\rho+q\right)}} +\partial_{r}\left(\rho+q\right)\overset{\sim\mathcal{O}\left(\delta^{2}\right)} {\overbrace{u_{r}\delta u^{r}}}\nonumber \\
 & +\left(\rho+q\right)\left(u^{t}\delta\left(\nabla_{t}u_{r}\right)+\delta u^{\nu}\nabla_{\nu}u_{r}\right)+k_{r}k^{r}\partial_{r}\delta\sigma\nonumber \\
 & +\delta\left(k_{r}k^{r}\right)\partial_{r}\sigma +k^{r}\left(\nabla_{r}k_{r}\right)\delta\sigma +\sigma\delta\left(\nabla_{r}k_{r}\right)k^{r}+\sigma\nabla_{\nu}k_{r}\delta k^{\nu}\nonumber \\
 & +\delta\sigma k_{r}\nabla_{\nu}k^{\nu}+\sigma\delta k_{r}\nabla_{\nu}k^{\nu}+\sigma k_{r}\delta\left(\nabla_{\nu}k^{\nu}\right)+\partial_{r}\delta q\nonumber \\
 & \;=\;\left(\delta\rho+\delta q\right)a_{r}+\left(\rho+q\right)u^{t}\left[\partial_{t}\delta u_{r}-\delta\Gamma_{tr}^{t}u_{t} -\overset{\sim\mathcal{O}\left(\delta^{2}\right)} {\overbrace{\Gamma_{tr}^{r}\delta u_{r}}}-\overset{\sim\mathcal{O}\left(\delta^{2}\right)} {\overbrace{\Gamma_{tr}^{\theta}\delta u_{\theta}}} -\overset{\sim\mathcal{O}\left(\delta^{2}\right)} {\overbrace{\Gamma_{tr}^{\phi}\delta u_{\phi}}}\right]
 \nonumber \\
 & +\left(\rho+q\right)\left[\overset{\sim\mathcal{O}\left(\delta^{2}\right)}{\overbrace{\delta u^{r}\partial_{r}u_{r}}}-\overset{\sim\mathcal{O}\left(\delta^{2}\right)}{\overbrace{\delta u^{r}\Gamma_{rr}^{t}}}u_{t}+\overset{\sim\mathcal{O}\left(\delta^{2}\right)}{\overbrace{\delta u^{\theta}\partial_{\theta}u_{r}}}-\overset{\sim\mathcal{O}\left(\delta^{2}\right)}{\overbrace{\delta u^{\theta}\Gamma_{r\theta}^{t}}}u_{t}+\overset{\sim\mathcal{O}\left(\delta^{2}\right)}{\overbrace{\delta u^{\phi}\partial_{\phi}u_{r}}}-\overset{\sim\mathcal{O}\left(\delta^{2}\right)}{\overbrace{\delta u^{\phi}\Gamma_{r\phi}^{t}}}u_{t}\right]\nonumber \\
 & +\partial_{r}\delta p+k^{r}\left(\partial_{r}k_{r}-\Gamma_{rr}^{r}k_{r}\right)\delta\sigma+\sigma\left(\partial_{r}\delta k_{r}-\delta\Gamma_{rr}^{r}k_{r}-\Gamma_{rr}^{\alpha}\delta k_{\alpha}\right)k^{r}\nonumber \\
 & +\sigma\left[\overset{\sim\mathcal{O}\left(\delta^{2}\right)}{\overbrace{\partial_{t}k_{r}\delta k^{t}}}-k_{r}\overset{\sim\mathcal{O}\left(\delta^{2}\right)}{\overbrace{\Gamma_{rt}^{r}\delta k^{t}}}+\left(\partial_{r}k_{r}-\Gamma_{rr}^{r}k_{r}\right)\delta k^{r}\right]\nonumber \\
 & +\sigma\left[\overset{\sim\mathcal{O}\left(\delta^{2}\right)}{\overbrace{\delta k^{\theta}\partial_{\theta}k_{r}-k_{r}\Gamma_{r\theta}^{r}\delta k^{\theta}}}+\overset{\sim\mathcal{O}\left(\delta^{2}\right)}{\overbrace{\delta k^{\phi}\partial_{\phi}k_{r}-k_{r}\Gamma_{r\phi}^{r}\delta k^{\phi}}}\right]\nonumber \\
 & +\left(\delta\sigma k_{r}+\sigma\delta k_{r}\right)\nabla_{\nu}k^{\nu}+\sigma k_{r}\left(\partial_{\nu}\delta k^{\nu}+\delta\Gamma_{r\nu}^{\nu}k^{r}+\Gamma_{\alpha\nu}^{\nu}\delta k^{\alpha}\right)\;,
\end{align}
or, more explicitly,
\begin{align}
0\;=\;\delta\left(T_{r\phantom{;\nu};\nu}^{\phantom{\nu}\nu}\right) & \;=\;\left(\delta\rho+\delta q\right)a_{r}+\left(\rho+q\right)u^{t}\left[\partial_{t}\delta u_{r}-\delta\Gamma_{tr}^{t}u_{t}\right]+\partial_{r}\delta p\nonumber \\
 & +k^{r}\left(\partial_{r}k_{r}-\Gamma_{rr}^{r}k_{r}\right)\delta\sigma+\sigma\left(\partial_{r}\delta k_{r}-\delta\Gamma_{rr}^{r}k_{r}-\Gamma_{rr}^{\alpha}\delta k_{\alpha}\right)k^{r}\nonumber \\
 & +\sigma\left(\partial_{r}k_{r}-\Gamma_{rr}^{r}k_{r}\right)\delta k^{r}+\left(\delta\sigma k_{r}+\sigma\delta k_{r}\right)\nabla_{\nu}k^{\nu}\nonumber \\
 & +\sigma k_{r}\left(\partial_{\nu}\delta k^{\nu}+\delta\Gamma_{r\nu}^{\nu}k^{r}+\Gamma_{\alpha\nu}^{\nu}\delta k^{\alpha}\right)\nonumber \\
 & \;=\;\left\{ \partial_{r}+\Phi^{\prime}\left(\frac{\partial\rho}{\partial p}+1\right)+\frac{2}{r}\left(1-\frac{\partial q}{\partial p}\right)\right\} \delta p\nonumber \\
 & +\omega^{2}e^{-2\Phi}\left(\rho+p\right)H_{1}e^{i\omega t}Y_{\ell m}-\omega^{2}e^{-2\Phi}\left(\rho+p\right)r^{-2}e^{\Lambda}We^{i\omega t}Y_{\ell m}\nonumber \\
 & +\frac{1}{2}\left(\rho+p\right)H_{0}^{\prime}e^{i\omega t}Y_{\ell m}-\sigma K^{\prime}e^{i\omega t}Y_{\ell m}\nonumber \\
 & +\sigma e^{\Lambda}\widehat{\delta k^{\theta}}\left[\overset{\;=\;-\ell\left(\ell+1\right)Y_{\ell m}}{\overbrace{\frac{1}{\sin\theta}\partial_{\theta}\left(\sin\theta\partial_{\theta}Y_{\ell m}\right)+\sin^{-2}\theta\partial_{\phi}^{2}Y_{\ell m}}}\right]\nonumber \\
 & \;=\;\left\{ \partial_{r}+\Phi^{\prime}\left(\frac{\partial\rho}{\partial p}+1\right)+\frac{2}{r}\left(1-\frac{\partial q}{\partial p}\right)\right\} \left[-\frac{dp}{d\rho}\left(\rho+p\right)\frac{e^{-\Lambda}W^{\prime}}{r^{2}}e^{i\omega t}Y_{\ell m}\right.\nonumber \\
 & -\frac{dp}{d\rho}\left(\rho+q\right)\frac{\ell\left(\ell+1\right)}{r^{2}}Ve^{i\omega t}Y_{\ell m}+\frac{dp}{d\rho}\frac{2\sigma}{r^{3}}e^{-\Lambda}We^{i\omega t}Y_{\ell m}\nonumber \\
 & \left.-\frac{dp}{dr}e^{-\Lambda}\frac{W}{r^{2}}e^{i\omega t}Y_{\ell m}+\frac{1}{2}\frac{dp}{d\rho}\sigma H_{0}e^{i\omega t}Y_{\ell m}+\frac{dp}{d\rho}\left(\rho+q\right)Ke^{i\omega t}Y_{\ell m}\right]\nonumber \\
 & +\omega^{2}e^{-2\Phi}\left(\rho+p\right)H_{1}e^{i\omega t}Y_{\ell m}-\omega^{2}e^{-2\Phi}\left(\rho+p\right)r^{-2}e^{\Lambda}We^{i\omega t}Y_{\ell m}\nonumber \\
 & +\frac{1}{2}\left(\rho+p\right)H_{0}^{\prime}e^{i\omega t}Y_{\ell m}-\sigma K^{\prime}e^{i\omega t}Y_{\ell m}\nonumber \\
 & +\sigma e^{\Lambda}\widehat{\delta k^{\theta}}\left[\overset{\;=\;-\ell\left(\ell+1\right)Y_{\ell m}}{\overbrace{\frac{1}{\sin\theta}\partial_{\theta}\left(\sin\theta\partial_{\theta}Y_{\ell m}\right)+\sin^{-2}\theta\partial_{\phi}^{2}Y_{\ell m}}}\right]\;,
\end{align}
where the  last step has incorporated the definition of  $\delta k^{\theta}$ from
Eq.~(\ref{eq:deltaGthetar}).

If the free index is instead $\theta$, then
\begin{align}
0\;=\;\delta\left(T_{\theta\phantom{;\nu};\nu}^{\phantom{\nu}\nu}\right) & \;=\;u^{\nu}\partial_{\nu}\left(\delta\rho+\delta q\right)u_{\theta}+\overset{\sim\mathcal{O}\left(\delta^{2}\right)}{\overbrace{\left(\delta\rho+\delta q\right)a_{\theta}}}\nonumber \\
 & +\partial_{\nu}\left(\rho+q\right)\delta\left(u_{\theta}u^{\nu}\right)+\left(\rho+q\right)\delta\left(\nabla_{\nu}u_{\theta}u^{\nu}\right)\nonumber \\
 & +\partial_{\nu}\delta\sigma k_{\theta}k^{\nu}+\partial_{\nu}\sigma\delta k_{\theta}k^{\nu}+\partial_{\nu}\sigma k_{\theta}\delta k^{\nu}\nonumber \\
 & +\delta\sigma\nabla_{\nu}k_{\theta}k^{\nu}+\sigma\delta\left(\nabla_{\nu}k_{\theta}\right)k^{\nu}+\sigma\nabla_{\nu}k_{\theta}\delta k^{\nu}\nonumber \\
 & +\delta\sigma k_{\theta}\nabla_{\nu}k^{\nu}+\sigma\delta k_{\theta}\nabla_{\nu}k^{\nu}+\sigma k_{\theta}\delta\left(\nabla_{\nu}k^{\nu}\right)+\partial_{\theta}\delta q\nonumber \\
 & \;=\;\overset{\;=\;0}{\overbrace{u^{t}\partial_{t}\left(\rho+q\right)}}\delta u_{\theta}+\left(\rho+q\right)\left(\delta\left(\nabla_{t}u_{\theta}\right)u^{t}+\nabla_{\nu}u_{\theta}\delta u^{\nu}\right)\nonumber \\
 & +\partial_{r}\sigma\delta k_{\theta}k^{r}+\delta\sigma\nabla_{r}k_{\theta}k^{r}+\sigma\delta\left(\nabla_{r}k_{\theta}\right)k^{r}+\sigma\nabla_{\nu}k_{\theta}\delta k^{\nu}\nonumber \\
 & +\sigma\delta k_{\theta}\nabla_{\nu}k^{\nu}+\partial_{\theta}\delta q\nonumber \\
 & \;=\;\left(\rho+q\right)u^{t}\left(\partial_{t}\delta u_{\theta}-\overset{\sim\mathcal{O}\left(\delta^{2}\right)}{\overbrace{\Gamma_{\theta t}^{r}\delta u_{r}}}-\overset{\sim\mathcal{O}\left(\delta^{2}\right)}{\overbrace{\Gamma_{\theta t}^{\theta}\delta u_{\theta}}}-\overset{\;=\;0}{\overbrace{\Gamma_{\theta t}^{\phi}}}\delta u_{\phi}-\delta\Gamma_{\theta t}^{t}u_{t}\right)\nonumber \\
 & +\left(\rho+q\right)\left[-u_{t}\overset{\sim\mathcal{O}\left(\delta^{2}\right)}{\overbrace{\Gamma_{\theta r}^{t}\delta u^{r}}}-u_{t}\overset{\sim\mathcal{O}\left(\delta^{2}\right)}{\overbrace{\Gamma_{\theta\theta}^{t}\delta u^{\theta}}}-\overset{\;=\;0}{\overbrace{\Gamma_{\theta\phi}^{t}}}u_{t}\delta u^{\phi}\right]+\partial_{r}\sigma\delta k_{\theta}k^{r}-\overset{\sim\mathcal{O}\left(\delta^{2}\right)}{\overbrace{\Gamma_{\theta r}^{r}\delta\sigma}}\nonumber \\
 & +\sigma\left(\partial_{r}\delta k_{\theta}-\delta\Gamma_{\theta r}^{r}k_{r}-\overset{\sim\mathcal{O}\left(\delta^{2}\right)}{\overbrace{\Gamma_{\theta r}^{t}\delta k_{t}}}-\overset{\sim\mathcal{O}\left(\delta^{2}\right)}{\overbrace{\Gamma_{\theta r}^{r}\delta k_{r}}}-\Gamma_{\theta r}^{\theta}\delta k_{\theta}-\overset{\;=\;0}{\overbrace{\Gamma_{\theta r}^{\phi}}}\delta k_{\phi}\right)k^{r}\nonumber \\
 & +\sigma\left[-k_{r}\overset{\sim\mathcal{O}\left(\delta^{2}\right)}{\overbrace{\Gamma_{\theta t}^{r}\delta k^{t}}}-k_{r}\overset{\sim\mathcal{O}\left(\delta^{2}\right)}{\overbrace{\Gamma_{\theta r}^{r}\delta k^{r}}}-\Gamma_{\theta\theta}^{r}k_{r}\delta k^{\theta}-\overset{\;=\;0}{\overbrace{\Gamma_{\theta\phi}^{r}}}k_{r}\delta k^{\phi}\right]\nonumber \\
 & +\sigma\delta k_{\theta}\left[\partial_{r}k^{r}+\Gamma_{r\nu}^{\nu}k^{r}\right]+\partial_{\theta}\delta q\;,
\end{align}
which can be somewhat simplified to
\begin{align}
0\;=\;\delta\left(T_{\theta\phantom{;\nu};\nu}^{\phantom{\nu}\nu}\right) & \;=\;\partial_{\theta}\delta q+\left(\rho+q\right)e^{-2\Phi}\omega^{2}Ve^{i\omega t}\partial_{\theta}Y_{\ell m}+\frac{1}{2}\left(\rho+p\right)H_{0}e^{i\omega t}\partial_{\theta}Y_{\ell m}\nonumber \\
 & +e^{-\Lambda}\sigma\partial_{r}\delta k_{\theta}+e^{-\Lambda}\sigma\delta k_{\theta}\left(\frac{2}{r}+\Phi^{\prime}+\partial_{r}\ln\sigma\right)\nonumber \\
 & \;=\;-\frac{dq}{d\rho}\left[\left(\rho+p\right)\frac{e^{-\Lambda}W^{\prime}}{r^{2}}+\left(\rho+q\right)\frac{\ell\left(\ell+1\right)}{r^{2}}V\right]\nonumber \\
 & +\frac{dq}{d\rho}\left(\frac{2\sigma}{r}-\frac{d\rho}{dr}\right)e^{-\Lambda}\frac{W}{r^{2}}+\frac{1}{2}\frac{dq}{d\rho}\sigma H_{0}+\frac{dq}{d\rho}\left(\rho+q\right)K\nonumber \\
 & +\left(\rho+q\right)e^{-2\Phi}\omega^{2}V+\frac{1}{2}\left(\rho+p\right)H_{0}\nonumber \\
 & +\frac{e^{-2\Lambda-2\Phi}}{2}\left\{ e^{2\Phi}H_{0}^{\prime\prime}-e^{2\Phi}K^{\prime\prime}-\left(\frac{\partial_{r}\sigma}{\sigma}+\Lambda^{\prime}+2\Phi^{\prime}\right)\omega^{2}H_{1}+\omega^{2}H_{1}^{\prime}\right.\nonumber \\
 & -e^{2\Phi}H_{0}\left[2\Phi^{\prime}\left(\Lambda^{\prime}-\frac{\partial_{r}\sigma}{\sigma}\right)-2\Phi^{\prime\prime}\right]\nonumber \\
 & \left.-\left[\Lambda^{\prime}-2\Phi^{\prime}+\frac{\partial_{r}\sigma}{\sigma}\right]e^{2\Phi}H_{0}^{\prime}+\left(\Lambda^{\prime}+\frac{\partial_{r}\sigma}{\sigma}\right)e^{2\Phi}K^{\prime}\right\} \nonumber \\
 & +\frac{e^{-2\Lambda-2\Phi}}{2}\left[\omega^{2}H_{1}+e^{2\Phi}\left(2\Phi^{\prime}H_{0}+H_{0}^{\prime}-K^{\prime}\right)\right]\left(\frac{2}{r}+\Phi^{\prime}+\partial_{r}\ln\sigma\right)\;,
\end{align}
where the  term involving $H_{1}^{\prime}$ has been eliminated by using
Eq.~(\ref{eq:H1'}).

\end{document}